\documentclass[a4paper,11pt]{article}
\pdfoutput=1 

\usepackage{jcappub} 
\usepackage[T1]{fontenc} 
\usepackage[utf8]{inputenc}
\usepackage{amsfonts}
\usepackage{lmodern}
\usepackage{amsthm,graphicx}
\usepackage{epsfig}
\usepackage[normalem]{ulem}
\usepackage{latexsym,amssymb} 
\usepackage{amsmath,mathtools}
\usepackage{xcolor,xspace}
\usepackage{siunitx}
\usepackage[font=footnotesize, labelfont=bf]{subcaption}
\usepackage{booktabs} 

\newcommand*{\diff}[1][]{\mathop{}\!\mathrm{d}^{#1}}
\newcommand*{\diffrac}[3][]{\frac{\diff[#1]#2}{\diff#3^{#1}}}

\newcommand{\PBH}{\text{PBH}}
\newcommand*{\emit}{\text{em}}

\newcommand*{\elec}{\text{e}}
\newcommand*{\phot}{\gamma}

\newcommand*{\hy}{\text{H}}
\newcommand*{\he}{\text{He}}
\newcommand*{\alphaB}{\alpha_\text{B}}
\newcommand*{\betaB}{\beta_\text{B}}
\newcommand*{\eff}{\text{eff}}
\newcommand{\boltz}{k_\text{B}}
\newcommand*{\heat}{\text{h}}
\newcommand*{\ion}{\text{i}}
\newcommand*{\thomp}{\sigma_\text{T}}
\newcommand*{\ar}{a_\text{r}}

\newcommand*{\rhoc}{\rho_\text{c}}
\newcommand*{\ns}{n_\text{s}}
\newcommand*{\mat}{\text{m}}
\newcommand*{\bary}{\text{b}}
\newcommand*{\cdm}{\text{cdm}}
\newcommand*{\rad}{\gamma}

\newcommand*{\Omegab}[1][]{\Omega_{\bary#1}}
\newcommand*{\Omegac}[1][]{\Omega_{\cdm#1}}

\newcommand*{\dep}{\text{dep}}
\newcommand*{\inj}{\text{inj}}
\renewcommand*{\max}{\text{max}}

\newcommand*{\Planck}{\emph{Planck}\xspace}
\newcommand*{\CLASS}{\textsc{class}\xspace}
\newcommand*{\RECFAST}{\textsc{recfast}\xspace}
\newcommand*{\HyRec}{\textsc{HyRec}\xspace}
\newcommand*{\MultiNest}{\textsc{MultiNest}\xspace}

\newcommand*{\LCDM}{\(\Lambda\)CDM\xspace}

\newcommand*{\ten}{{10}}

\usepackage{color}

\DeclareSIUnit\Msol{\mathnormal{M_\odot}}

\graphicspath{{figs/}}

\title{CMB constraints on ultra-light primordial black holes with extended mass distributions}

\author[a]{Harry Poulter,}
\author[b]{Yacine Ali-Ha\"imoud,}
\author[c]{Jan Hamann,}
\author[a]{Martin White,}
\author[a]{Anthony G.~Williams}

\affiliation[a]{ARC Centre of Excellence for Particle Physics at the Terascale (CoEPP) \& CSSM, Department of Physics, University of Adelaide, South Australia 5005, Australia}
\affiliation[b]{Center for Cosmology and Particle Physics, Department of Physics, New York University, New York, NY 10003, USA}
\affiliation[c]{School of Physics, The University of New South Wales, Sydney NSW 2052, Australia}

\emailAdd{harry.poulter@adelaide.edu.au}
\emailAdd{yah2@nyu.edu}
\emailAdd{jan.hamann@unsw.edu.au}
\emailAdd{martin.white@adelaide.edu.au}
\emailAdd{anthony.williams@adelaide.edu.au}

\abstract{
We examine the effects ultra-light primordial black holes (PBHs) have on the anisotropies of the cosmic microwave background (CMB).
PBHs in the mass range of \SIrange{e15}{e17}{\gram} emit Hawking radiation in the early Universe, modifying the standard recombination history.
This leads to a damping of small-scale temperature and polarisation anisotropies and enhances large-scale polarisation fluctuations.
As some models of inflation predict PBHs with a range of masses, we investigate the impacts of extended mass distributions on PBH abundance constraints.
We model PBH energy injection using a ground-up approach incorporating species-dependent deposition efficiencies.
By allowing the \LCDM parameters to vary simultaneously with the PBH fraction and mass, we show that exclusion bounds on the PBH fraction of DM \(f_\PBH\) are relaxed by up to an order of magnitude, compared to the case of fixed \LCDM parameters.
We also give \SI{95}{\percent} exclusion regions for \(f_\PBH\) for a variety of mass distributions.
In particular, for a uniform mass distribution between \num{e15} and \SI{e17}{\gram}, we find \(f_\PBH < \num{1.6e-5}\) when allowing \LCDM parameters to vary.
}

\begin{document}
\maketitle

\defcitealias{Clark:2016nst}{CDG17}

\section{Introduction \label{sec:intro}}

The cosmological effects of primordial black holes (PBHs) were first considered by Chapline~\cite{Chapline:1975} in the mid '70s.
Originally proposed to have formed through collapse of overdense regions during the early universe~\cite{Hawking:1971ei}, PBHs have been considered a possible dark matter (DM) candidate for quite some time due to their invisibility in the electromagnetic spectrum and massive nature~\cite{Carr:2016drx}.
The conceptually viable range of dark matter PBH masses spans about thirty orders of magnitude, from \SI{e15}{\gram} to \SI{e12}{\Msol} --- any lighter, and such PBHs would have evaporated by today.
Any heavier and a single PBH would exceed the typical galactic halo mass, effectively ruling out their possible contribution to galactic dark matter.

Astrophysical observations such as microlensing~\cite{Niikura:2017zjd}, neutron star capture~\cite{Capela:2013yf}, large-scale structure considerations~\cite{Afshordi:2003zb}, CMB measurements~\cite{Ricotti:2008, Carr:2009jm,Bernal:2017vvn,AliHaimoud:2017,Poulin:2017a} and ultra-faint dwarf galaxies~\cite{Zhu:2017plg} have constrained large portions of this mass range.
Carr et al.~\cite{Carr:2016drx} identify three remaining mass windows of interest: intermediate mass PBHs of \(M_\PBH \sim\) \SIrange{10}{e2}{\Msol} (see however arguments based on LIGO limits on the binary-black-hole merger rate~\cite{AliHaimoud:2017b}), sublunar PBHs with \mbox{\(\SI{e20}{\gram} \lesssim M_\mathrm{PBH} \lesssim \SI{e22}{\gram}\)}, and subatomic PBHs with \(M_\PBH \sim \SI{e17}{\gram}\).
In addition, earlier constraints based on femtolensing of gamma-ray bursts were recently criticized by Katz et al.~\cite{Katz:2018zrn}, so there may be another allowed window at \(\SI{e17}{\gram} \lesssim M_\mathrm{PBH} \lesssim \SI{e19}{\gram}\).

In this paper we consider light PBHs in the range of \SIrange{e15}{e17}{\gram}, which eject energy through Hawking radiation.
Outside of cosmological considerations, this mass range is already well constrained from diffuse gamma ray background measurements~\cite{Carr:2016hva,Fermi-LAT:2018pfs} and femtolensing surveys~\cite{Barnacka:2012bm}.
Current constraints from CMB measurements lie in close proximity to gamma ray background constraints~\cite{Carr:2009jm,Poulin:2016anj}, and up until recently have typically assumed a monochromatic mass distribution.
Although the formation of PBHs with a quasi monochromatic distribution is possible~\cite{Pi:2017gih}, other models of inflation predict extended mass distributions~\cite{Clesse:2015wea}.
Such distributions have been explored in this context through converting previous monochromatic constraints to extended mass distribution constraints~\cite{Carr:2017jsz,Kuhnel:2017pwq,Bellomo:2017zsr}.
This work takes a complementary approach by incorporating extended mass distributions into the model explicitly.
We note the multitude of different approaches taken in modelling Hawking radiation from light PBHs~\cite{Poulin:2016anj,Belotsky:2014twa,Clark:2016nst}, and choose to follow the method outlined in Ref.~\cite{Clark:2016nst}, hereafter~\citetalias{Clark:2016nst}.
This method lends itself well to extended mass distributions due to it being computationally quick to calculate, and in the case of monochromatic distributions gives results comparable to the more robust treatment given in Ref.~\cite{Poulin:2016anj}.

In Section~\ref{sec:theory} we introduce the dynamics behind evaporating PBHs, covering their ejection spectrum, particle content and energy injection rate.
We also recount the extension of the standard recombination model to incorporate non-standard energy injections.
Section~\ref{sec:method} outlines how the theory was implemented, including details on integration of the mass distribution and computation of deposition efficiencies.
The results are presented in Section~\ref{sec:results}.
Finally, Section~\ref{sec:conclusions} concludes the work, presenting possible future extensions.

Throughout this work, we work in units where \(\boltz = 1\).

\section{Light PBHs and the CMB \label{sec:theory}}

\subsection{Evaporation Physics}

PBHs in the mass range of \SIrange{e15}{e17}{\gram} inject energy into the primordial plasma through Hawking radiation.
In general, this process depends on the hole's temperature, angular momentum and charge.
For primordial black holes, however, we are able to simplify matters.
As the radiation contains particles that carry spin, PBHs lose angular momentum more quickly than they lose mass~\cite{Page:1976ki}.
We also assume that the formation mechanism responsible for PBHs results in charge-neutral holes.
Hence, at the recombination epoch we take PBHs to be neutral with no spin, which gives a direct relationship between their mass \(M_\PBH\) and temperature, as predicted by Hawking~\cite{Hawking:1974rv,Hawking:1974sw}:
\begin{align}
  \label{eq:pbh-temp}
  T_\PBH = \frac{\hbar c^3}{8\pi G M_\PBH} \approx \SI{1.1}{\mega\electronvolt}~ \frac{\SI{e16}{\gram}}{M_\PBH}\,.
\end{align}
Such a PBH emits particles with an energy spectrum given by the corresponding blackbody spectrum with temperature \(T_\PBH\).
Particles \(i\) (\(\elec^\pm\), \(\gamma\) and \(\nu, \overline{\nu}\)) are ejected with the following rate per particle degree of freedom, per energy interval:
\begin{equation}
  \label{eq:pbh-spectra}
  \frac{\diff N^i}{\diff t \diff E} = \frac{\Gamma_{s_i}(M_\PBH, E)}{2\pi\hbar}\,\frac{1}{\exp\left(\frac{E}{T_\PBH}\right) - (-1)^{2s_i}}\,.
\end{equation}
In this equation, \(s_i\) is the particle's spin, \(E\) is the total energy of the particle (including rest mass) and \(\Gamma_{s_i}(M_\PBH, E)\) is its associated dimensionless absorption probability, given by~\cite{MacGibbon:1990zk}
\begin{equation}
  \label{eq:pbh-absorb-prob}
  \Gamma_{s_i}(M_\PBH, E) = \frac{E^2 \sigma_{s_i}(M_\PBH, E)}{\pi \hbar^2 c^2}\,,
\end{equation}
where \(\sigma_{s_i}\) is the absorption cross section.
For high energies \(E \gg T_\PBH\), this probability approaches the optical limit \(\Gamma_{s_i}(M_\PBH, E) \to 27\,G^2 M_\PBH^2 E^2/\hbar^2 c^6\).
For low energies \(E \to 0\), we find that~\cite{MacGibbon:1990zk}
\begin{align}
  \label{eq:gamma-low-energy}
  \Gamma_{s_i}(M_\PBH, E) \sim
  \begin{dcases}
    \frac{2G^2M_\PBH^2E^2}{\hbar^2c^6}\,, & s_i = \tfrac{1}{2}\,, \\
    \frac{64G^4M_\PBH^4E^4}{3\hbar^4c^{12}}\,, & s_i = 1\,.
  \end{dcases}
\end{align}
Between these two limits, a numerical form is given in Figure~1 of Ref.~\cite{MacGibbon:1990zk}, which was later used to construct an approximation to the full \(\Gamma_{s_i}(M_\PBH, E)\).

The rate of mass lost by the PBH through Hawking radiation is given by~\cite{Carr:2009jm}:
\begin{align}
  \label{eq:pbh-evap}
  \diffrac{M_\PBH}{t} &= - \sum_i g_i \int \frac{\diff N^i}{\diff t \diff E} \frac{E}{c^2}\diff E \nonumber \\
                      &= - \num{5.34e-7} \times \left[\sum_i g_i f^\emit_i(M_\PBH)\right] \left(\frac{\SI{e16}{\gram}}{M_\PBH}\right)^2 ~\si{\gram\per\second}\,,
\end{align}
where \(g_i = 2 s_i +1\) are degeneracy factors for each particle and \(f_i^\emit(M_\PBH)\) are emission fractions.
The sum over \(f_i^\emit(M_\PBH)\) is normalised to unity for a PBH with mass \(M_\PBH \gg \SI{e17}{\gram}\), which would only emit photons and neutrinos as such black holes have temperatures \(T_\PBH \ll m_{\elec^{\pm}}\).

The contributions of photons and neutrinos (which are always relativistic for the PBH masses of interest) are independent of PBH mass and given by~\cite{MacGibbon:1991tj}
\begin{align*}
    f_\phot^\emit &= 0.060\,, &
    f_\nu^\emit &= 0.147\,.
\end{align*}

However, we need to be cautious with electrons and positrons at the high-end of our mass range \(M_\PBH \sim 10^{17}\) g, where \(T_\PBH \sim m_{\elec^\pm}\).
By reformulating Eq.~(7) from Ref.~\cite{MacGibbon:1991tj}, we get:
\begin{equation}
\label{eq:elec-emission}
    f_{\elec^\pm}^\emit(M_\PBH) = 0.142\, \exp\left(-\frac{M_\PBH}{M_{\elec^\pm}}\right), \qquad
    M_{\elec^{\pm}} \equiv \SI{9.4e16}{\gram}\,.
\end{equation}
In the limit \(M_\PBH \ll \SI{e17}{\gram}\), we find the above \(\sum_i g_i f_i^{\rm em} \rightarrow 1.569\), in agreement with the constant term in Eq.~(7) of Ref.~\cite{MacGibbon:1991tj}.

In the PBH mass range of \SIrange{e15}{e17}{\gram}, the mass lost from particle emission is negligible compared to the total PBH mass, and so we do not need to calculate the full expression given in Eq.~\eqref{eq:pbh-evap} for all emitted species.
Instead, we are more concerned about what effects these emitted particles may have on the early universe.
The particles emitted from PBHs in this mass range that interact most strongly with the primordial plasma are electrons, positrons and photons.
We define the mass lost through purely electromagnetic particles, \(\diff M_\PBH/\!\diff t|_\text{EM}\), as in Eq.~\eqref{eq:pbh-evap}, where the sum only runs over \(i\in \{\elec^-,\elec^+,\gamma\} \).
The total energy injected into the primordial plasma by evaporating PBHs is then
\begin{equation}
  \left(\frac{\diff E}{\diff V\diff t}\right)_\PBH = -\left.\diffrac{M_\PBH}{t}\right|_\text{EM}\, c^2 \times n_\PBH(z)\,,
\end{equation}
where \(n_\PBH(z) \) is the number density of PBHs.
As previously discussed, for the range of masses considered the mass and abundance of evaporating PBHs is essentially constant throughout cosmic history.
The number density can be written in terms of its density parameter \(\Omega_\PBH(z) = (1+z) \, \Omega_{\PBH,0}\), giving
\begin{equation*}
  n_\PBH(z) = \frac{\rhoc\Omega_\PBH(z)}{M_\PBH} = \frac{\rhoc \Omegac[,0] f_\PBH }{M_\PBH} (1+z)^3\,,
\end{equation*}
where we have defined the fraction of dark matter consisting of PBHs
\begin{equation*}
  f_\PBH = \frac{\Omega_{\PBH,0}}{\Omegac[,0]}\,,
\end{equation*}
and \(\rhoc= 3H_0^2/8\pi G\) is the critical density of the Universe today.
Hence, the volumetric rate of energy injection in the form of electrons, positrons and photons is
\begin{equation}
  \label{eq:pbh-energy-inj}
    \left(\frac{\diff E}{\diff V\diff t}\right)_\PBH = \num{5.34e-23} \times \left[\sum_{i= \gamma, e^{\pm}} g_i f^\emit_i(M_\PBH)\right] \left(\frac{\SI{e16}{\gram}}{M_\PBH}\right)^3 c^2 \rhoc \Omegac[,0] f_\PBH (1+z)^3 ~\si{\per\second}\,.
\end{equation}

\subsection{Effects on Recombination}

In order to examine the effects evaporating PBHs have on CMB anisotropies, the standard formulation of recombination needs to be modified to incorporate the extra energy injection.

In more general terms, if some non-standard process \(X\) injects energy into the ionised medium at some rate per unit comoving volume \((\diff E/\!\diff V\diff t)_\inj\), this can excite and ionise both hydrogen and helium atoms, heat the plasma and inject continuum photons with \(E < \SI{10.2}{\electronvolt}\) that contribute directly to CMB spectral distortions.
This leads to the definition of effective deposition efficiencies \(f_c(z)\) for each of these deposition channels, which encapsulate how injected energy is then deposited into some channel \(c\):
\begin{equation*}
  \left(\frac{\diff E}{\diff V\diff t}\right)_{c,\text{dep}} = f_{X,c}(z)\left(\frac{\diff E}{\diff V\diff t}\right)_\text{inj}\,.
\end{equation*}
In practice, however, we only consider the ionisation and excitation of hydrogen atoms, as helium recombines at a much earlier time and has a smaller effect on the CMB.
The amplitude of CMB spectral distortions is proportional to \(\int (\diff E/\diff V \diff t)/H \rho_\gamma \diff \ln(1+z)\) (see, e.g. \cite{Chluba_13}), and it is straightforward to show that it is always negligibly small.
We therefore focus on the effect on recombination history and CMB anisotropies.

In the standard three-level atom model, equations for the ionisation history and matter temperature are given by~\cite{Peebles:1968ja,Zeldovich:1969en}
\begin{align*}
  \diffrac{x_\elec}{z} &= \frac{C(x_\elec, T_\mat; z)}{(1+z)H(z)}\left\{\alphaB(T_\mat, T_\rad)\,x_\elec^2 n_\hy(z) - \betaB(T_\rad)\,(1-x_\elec) e^{-h\nu_\text{2s}/T_\mat}\right\}\,, \\
  \diffrac{T_\mat}{z} &= \frac{1}{(1+z)H(z)}\left\{\frac{8\thomp\ar T_\rad^4(z)}{3 m_\elec c^2}\,\frac{x_\elec}{1 + f_\he + x_\elec}\,[T_\mat - T_\rad(z)] + 2H(z)T_\mat\right\}\,,
\end{align*}
where \(C(x_\elec, T_\mat; z)\) is Peebles' coefficient, \(\alphaB(T_\mat, T_\rad)\) and \(\betaB(T_\rad)\) are free-bound recombination and ionisation rates for hydrogen, \(\nu_\text{2s}\) is the \(2s\to1s\) transition frequency for hydrogen, and \(f_\he\) is the ratio of helium nuclei to hydrogen nuclei.

For energy injected from a generic process, extra non-standard terms are given by~\cite{Chen:2003gz}
\begin{align*}
  \left(\diffrac{x_\elec}{z}\right)_\text{non std.} &= \frac{1}{(1+z)H(z)}\left[I_{X_\ion}(x_\elec, T_\mat; z) + I_{X_\alpha}(x_\elec, T_\mat; z)\right]\,, \\
    \left(\diffrac{T_\mat}{z}\right)_\text{non std.} &= \frac{2}{3(1+z)H(z)}\,\frac{K_\heat(x_\elec, T_\mat; z)}{1+f_\he+x_\elec}\,,
\end{align*}
where \(I_{X_\ion}\), \(I_{X_\alpha}\) and \(K_\heat\) are the ionisation, excitation and heating terms given by
\begin{align}
  I_{X_\ion}(x_\elec, T_\mat; z) &= f_{X,\ion}(z)\, \frac{(\diff E/\!\diff V\diff t)_\text{inj}}{n_\hy(z)E_\ion}\,, \label{eq:non-ion} \\
  I_{X_\alpha}(x_\elec, T_\mat; z) &= [1-C(x_\elec, T_\mat;z)] f_{X,\alpha}(z)\, \frac{(\diff E/\!\diff V\diff t)_\text{inj}}{n_\hy(z)E_\alpha}\,, \label{eq:non-excite} \\
  K_\heat(x_\elec, T_\mat; z) &= f_{X,\heat}(z)\, \frac{(\diff E/\!\diff V\diff t)_\text{inj}}{n_\hy(z)}\,. \label{eq:non-heat}
\end{align}

Here, \(n_\hy(z)\) is the number density of hydrogen nuclei, and \(E_\ion\) and \(E_\alpha\) are the ionisation and Lyman-\(\alpha\) energies for the hydrogen atom.
All three of these terms also rely on the injected energy and their related deposition efficiencies.
As their emitted radiation comprises of particles, PBH deposition efficiencies can in general be calculated by computing
\begin{equation}
  f_{\PBH,c}(M_\PBH, z) = \frac{\sum_i g_i f_i^\emit(M_\PBH) f_{\eff,c}^i(E_i, z)}{\sum_i g_i f_i^\emit(M_\PBH)}\,,
\end{equation}
which weights each emitted particle's effective deposition efficiency \(f_{\eff,c}^i(E_i, z)\) by its emission fraction \(f_i^\emit(M_\PBH)\).
Here, the effective deposition efficiencies are calculated from their corresponding differential energy absorption rates given by Slatyer~\cite{Slatyer:2015kla}.
We will describe their computation in more detail below.
For PBHs in the mass range of \SIrange{e15}{e17}{\gram}, the emitted particles relevant to recombination physics are electrons/positrons and photons.
As we have already seen, \(m_{\elec^\pm} \sim T_\PBH\) for PBHs with mass \SI{e17}{\gram}, so we use Eq.~\eqref{eq:elec-emission} for our form of \(f_{\elec^\pm}^\emit(M_\PBH)\) to model the turning-on of relativistic effects.
Hence, the net deposition efficiency is
\begin{equation}
  \label{eq:pbh-dep-eff}
  f_{\PBH,c}(M_\PBH, z) = \frac{4f^\emit_{\elec^\pm}(M_\PBH) f^{\elec^\pm}_{\eff,c}(E_\elec, z) + 2f^\emit_\gamma f^\gamma_{\eff,c}(E_\gamma,z)}{4 f^\emit_{\elec^\pm}(M_\PBH) + 2f^\emit_\gamma}\,.
\end{equation}

\subsection{Deposition Efficiencies}
Rather than using the common on-the-spot approximation, which assumes that all the injected energy is immediately deposited into the plasma, we compute the deposition efficiency from first principles.
The effective deposition efficiency for a particle species \(i\) is defined as~\cite{Slatyer:2015kla}
\begin{align}
  f^i_{\eff,c}(z_\dep) &= \frac{H(z_\dep)(1+z_\dep)^3}{\int E\left(\frac{\diff N^i}{\diff E\diff V\diff t}\right)_{z_\dep}\diff E} \nonumber \\
  &\qquad \times \iint E \left(\frac{\diff N^i}{\diff E\diff V\diff t}\right)_{z_\inj}\,\frac{T^i_c(z_\dep, E, z_\inj)}{H(z_\inj)(1+z_\inj)^3}\diff\ln(1+z_\inj)\diff E\,, \label{eq:fi-general}
\end{align}
where \(T^i_c(z_\dep, E, z_\inj)\) is the differential rate of energy absorption at \(z_\dep\) for a particle injected at \(z_\inj\) with total energy \(E\) and
\begin{equation*}
  \left(\frac{\diff N^i}{\diff E\diff V\diff t}\right)_z = n_\PBH(z)\,\frac{\diff N^i}{\diff t \diff E} \propto (1 + z)^3 \frac{\diff N^i}{\diff t \diff E}\,,
\end{equation*}
where \(\diff N^i/\diff t \diff E\) is given by Eq.~\eqref{eq:pbh-spectra}.

In principle, due to its dependency on the Hubble rate, our \(f^i_{\eff,c}(M_\PBH, z_\dep)\) should be calculated for each point in \LCDM parameter space.
However, as we are focusing on the region of \(1 \ll z \ll 3000\) deep in the matter-dominated era, the Hubble parameter can be approximated by \(H(z) \approx H_0 (1+z)^{3/2}\).
Along with the above form of the ejection spectrum, the effective deposition efficiencies simplify to
\begin{align}
    \label{eq:eff-dep}
  f^i_{\eff,c}(M_\PBH, z_\dep) &\approx \frac{(1+z_\dep)^{3/2}}{\int E\,\frac{\diff N^i}{\diff t\diff E}\diff E} \iint E\, \frac{\diff N^i}{\diff t\diff E}\,\frac{T^i_c(z_\dep, E, z_\inj)}{(1+z_\inj)^{3/2}} \diff\ln(1+z_\inj)\diff E.
\end{align}
As this quantity is now independent of any cosmological parameters, it can be computed ahead of time and stored in a lookup table for later use.

\subsection{Extended Mass Distributions}

We now reformulate the non-standard energy injection problem to account for an extended PBH mass distribution.
Instead of considering energy being injected by PBHs at a single mass, we must integrate over a range of masses weighted by a probability distribution \(p(M_\PBH|\vec{\theta})\) dependent on shape parameters \(\vec{\theta}\).
For definiteness, we define \(p(M_\PBH|\vec{\theta})\) to be the probability density per \(\log_{10}\) mass interval.
This leads us to define the following quantity:
\begin{align}
  F_{\PBH,c}(z;\vec{\theta}) :\!&= \int p(M_\PBH|\vec{\theta}) \left(\frac{\diff E}{\diff V\diff t}\right)_{c,\text{dep}} \diff\log_\ten \left(M_\PBH\right) \nonumber \\
  &= \int p(M_\PBH|\vec{\theta})\, f_{\PBH,c}(M_\PBH, z)\left(\frac{\diff E}{\diff V\diff t}\right)_\text{inj} \diff\log_\ten \left(M_\PBH\right)\,. \label{eq:F-frac}
\end{align}

This quantity is then used in the non-standard ionisation, excitation and heating terms of Eqs.~\eqref{eq:non-ion}-\eqref{eq:non-heat}:
\begin{align*}
    I_{X_\ion}(x_\elec, T_\mat; z) &= \frac{F_{\PBH,\ion}(z;\vec{\theta})}{n_\hy(z)E_\ion}\,, \\
  I_{X_\alpha}(x_\elec, T_\mat; z) &= [1-C(x_\elec, T_\mat;z)] \frac{F_{\PBH,\alpha}(z;\vec{\theta})}{n_\hy(z)E_\alpha}\,, \\
  K_\heat(x_\elec, T_\mat; z) &= \frac{F_{\PBH,\heat}(z;\vec{\theta})}{n_\hy(z)}\,. 
\end{align*}

In this work we choose a log-normal distribution for our \(p(M_\PBH|\vec{\theta})\), which in part is motivated by the large perturbations that will eventually collapse into PBHs being generated during the inflationary phase of the Universe.
In this case, the scale factor \(a \propto \exp (H t)\), leading to physical wavenumbers \(k \propto a^{-1}\).
Therefore, if whatever process that leads to the formation of large density perturbations occurs on a timescale \(\Delta t\) during inflation, the natural scale of the modulation of the primordial power spectrum is proportional to \(\Delta \ln k\), which corresponds to \(\Delta \ln M\) in the PBH mass function.
The log-normal distribution has also been examined in other works~\cite{Green:2016xgy,Kannike:2017bxn}.

The mass distribution we use takes the form
\begin{equation}
  p(M_\PBH|\mu_\ten, \sigma_\ten) = \frac{1}{\sqrt{2\pi} \, \sigma_\ten}\,\exp\left[-\frac{\left(\log_\ten(M_\PBH) - \mu_\ten\right)^2}{2 \, \sigma_\ten^2}\right]\,,
  \label{eq:log-norm-dist}
\end{equation}
where \(\mu_\ten\) is the mean and \(\sigma_\ten\) is the standard deviation (or effective width) of the distribution.
As we are working in \(\log_\ten\) space, \(\sigma_\ten\) denotes how many orders of magnitude the mass distribution spans.

In this work we consider a normal distribution in log space, equivalent to a log-normal in linear space as considered in other works.
We choose this form of the distribution over the more familiar one presented in the literature as its symmetry allows it to be integrated easily using a trapezoidal method in log space.
This choice of distribution can be viewed as identical to those considered in~\cite{Carr:2017jsz} under the following transformations:
\begin{align*}
  \sigma &= \sigma_\ten \ln(10)\,, &
  M_c &= 10^{16 + \mu_\ten}\,.
\end{align*}

\subsection{Extending Gamma Ray Background Constraints}
\label{subsec:extending-constraints}

CMB constraints for evaporating PBHs are competing with limits derived from gamma ray background measurements.
We would like to compare these constraints for extended mass distributions.
For this, we use the method outlined in Ref.~\cite{Carr:2017jsz} which we summarise here.

Consider an astrophysical observable \(A[\psi(M_\PBH)]\) dependent on the PBH mass function \(\psi(M_\PBH) = f_\PBH \times p(M_\PBH|\vec{\theta})\), where again \(p(M_\PBH|\vec{\theta})\) is the probability distribution per \(\log_{10}\) mass interval, and \(f_{\PBH}\) is the total PBH fraction.
The observable can be expanded:
\begin{align*}
  A[\psi(M_\PBH)] &= A_0 + f_\PBH \int K_1(M_\PBH) p(M_\PBH|\vec{\theta}) \diff \log_\ten(M_\PBH) \\
  &\quad {} + f_\PBH^2\int K_2(M_1,M_2)\ p(M_1|\vec{\theta}) p(M_2|\vec{\theta}) \diff \log_\ten(M_1)\diff \log_\ten(M_2) + \dotsb\,,
\end{align*}
for some background contribution \(A_0\) and observation-dependent functions \(K_j\).
As we are dealing with evaporating PBHs whose contribution at a certain mass is independent of PBHs of different masses, all \(K_j = 0\) for \(j > 1\) and we only consider the first two terms in the sum.
Suppose the measurement places an upper bound on this observable such that
\begin{equation*}
  A-A_0 \leq A_\max\,.
\end{equation*}
For a monochromatic distribution centered at mass \(M_c\), we have
\begin{equation*}
  p_\text{mono}(M_\PBH|M_c) = \delta\left[\log_{10}(M_\PBH) - \log_{10}(M_c)\right]\,,
\end{equation*}
implying an upper limit to the PBH fraction
\begin{equation*}
  f_{\PBH,\text{mono}}(M_c) \leq \frac{A_\max}{K_1(M_c)} := f_\text{max}(M_c)\,.
\end{equation*}
This allows us to infer \(K_1(M) = A_{\max}/f_{\max}(M)\).
Hence the upper limit on the PBH fraction for a general distribution \(p(M_\PBH|\vec{\theta})\) is
\begin{equation*}
  f_{\PBH}(\vec{\theta}) \leq  \left(\int \frac{p(M_\PBH|\vec{\theta})}{f_\text{max}(M_\PBH)} \diff \log_\ten(M_\PBH) \right)^{-1}\,.
\end{equation*}

For this work, we use the following constraint from gamma ray background observations~\cite{Carr:2016drx}:
\begin{align}
  f_\text{max}(M_\PBH) &= \num{2e-8} \left(\frac{M_\PBH}{\SI{5e14}{\gram}}\right)^{3+\epsilon}\,, &
  M &< \SI{e18}{\gram}\,, \label{eq:pbh-frac-gamma-mono}
\end{align}
where \(\epsilon\) characterises the spectral tilt in the gamma ray background, set to \(\epsilon = 0.4\) here.
This relationship constrains PBHs that emit gamma radiation solely through primary photons, which is true for PBHs with \(M_\PBH \gtrsim \SI{5e14}{\gram}\).

\section{Data and Methodology \label{sec:method}}

\subsection{Efficiency Computation}

The PBH deposition efficiencies were calculated using the particle deposition efficiencies for electrons/positrons and photons.
These were in turn computed using Eq.~\eqref{eq:eff-dep} alongside the differential energy absorption rates \(T^i_c(z_\dep, E, z_\inj)\) given in Ref.~\cite{Slatyer:2015kla} and the forms of \(\Gamma_{s_i}(M_\PBH,E)\) given in Ref.~\cite{MacGibbon:1990zk}.

In this work, a piecewise function was created for \(\Gamma_{s_i}(M_\PBH,E)\) from interpolating the numerical form given in Ref.~\cite{MacGibbon:1990zk} (indicated by \(\Gamma^\text{num.}_{s_i}(M_\PBH, E)\)) and using the low-energy limit from Eq.~\eqref{eq:gamma-low-energy}.
Due to the range of energies and PBH masses considered, the high-energy optical limit was not required.

\begin{figure}[!t]
  \centering
  \begin{subfigure}[t]{0.49\textwidth}
    \centering
    \includegraphics[height=2.4in]{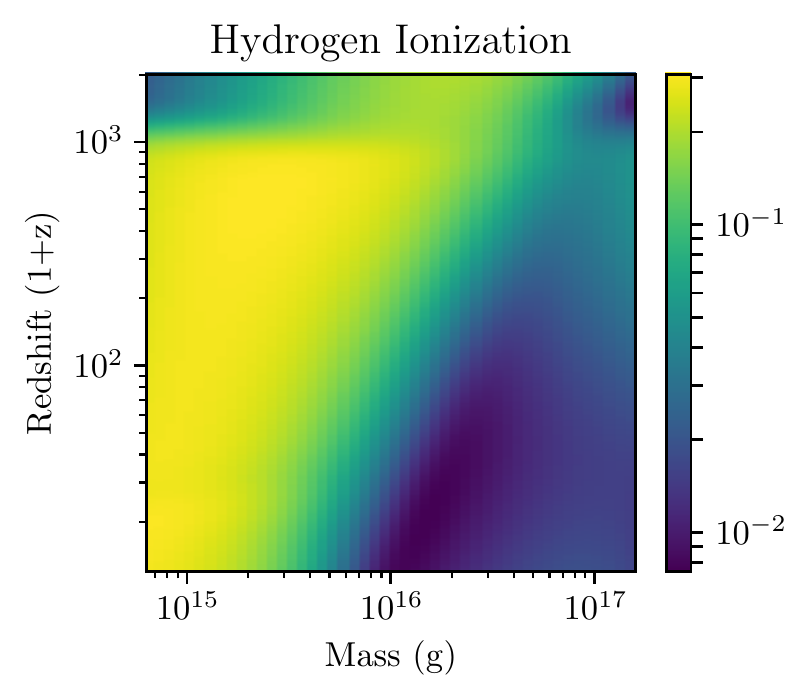}
    \label{subfig:pbh-hion}
  \end{subfigure}%
  ~
  \begin{subfigure}[t]{0.49\textwidth}
    \centering
    \includegraphics[height=2.4in]{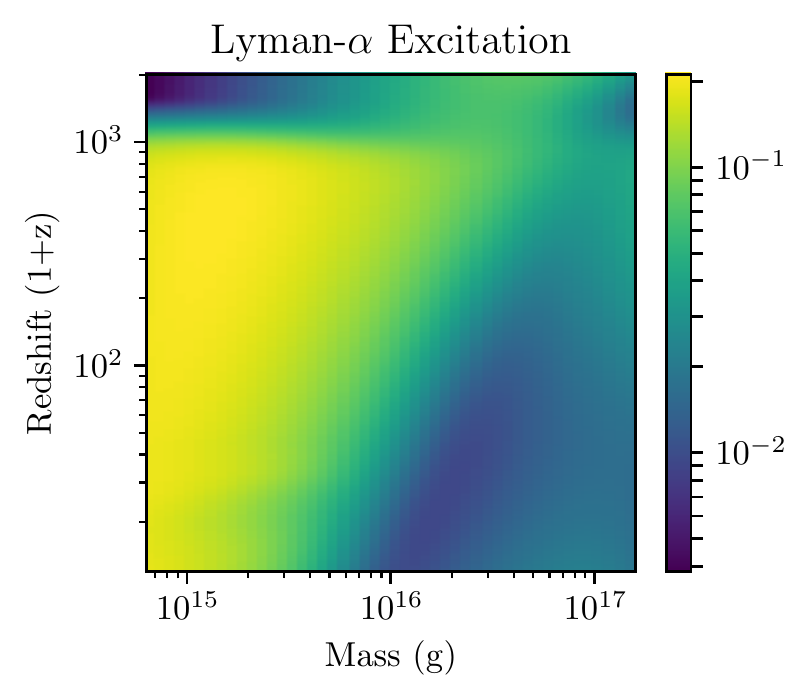}
    \label{subfig:pbh-excite}
  \end{subfigure}
  \begin{subfigure}[t]{0.49\textwidth}
    \centering
    \vspace{-1.0em}
    \includegraphics[height=2.4in]{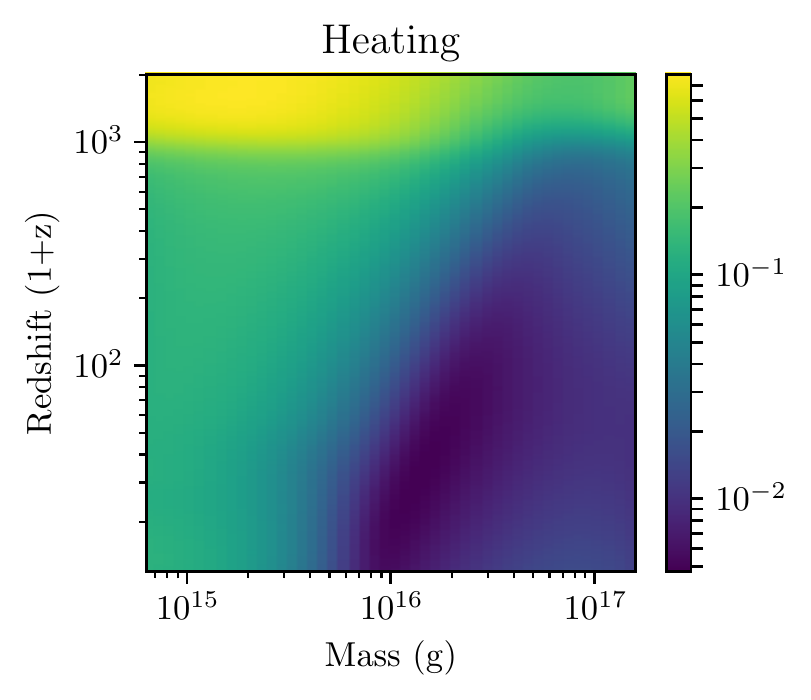}
    \label{subfig:pbh-heat}
  \end{subfigure}%
  ~
  \begin{subfigure}[t]{0.49\textwidth}
    \centering
    \vspace{-1.0em}
    \includegraphics[height=2.4in]{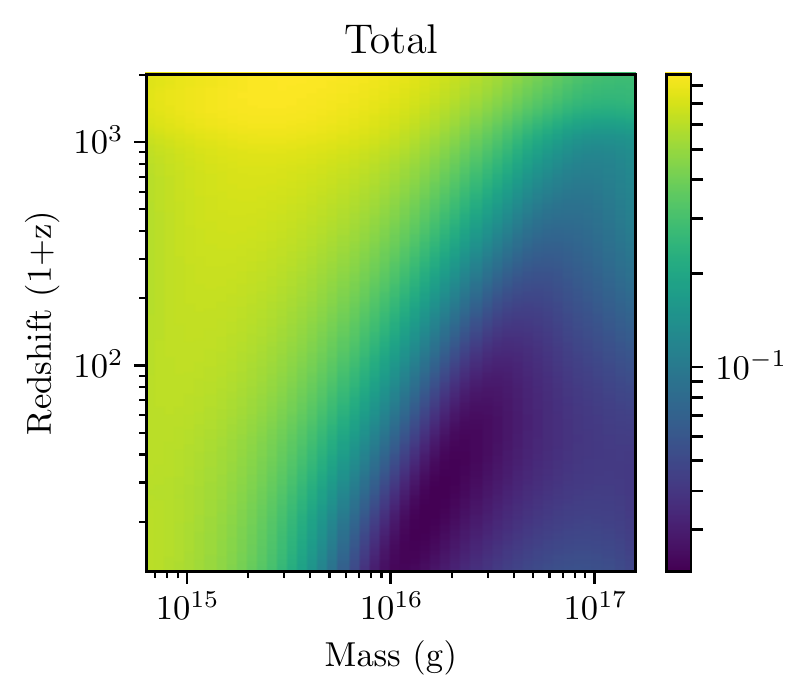}
    \label{subfig:5_pbh-total}
  \end{subfigure}
  \vspace{-1.0em}
  \caption{Energy deposition efficiencies \(f_{\PBH,c}(M_\PBH,z)\). The temperature axis in Figure~1 of~\citetalias{Clark:2016nst} has been transformed into a mass axis in this work. The total is a sum over the three displayed channels only.}
  \label{fig:pbh-fdep}
\end{figure}

For electrons, this took the form of
\begin{equation*}
    \Gamma_{1/2}(M_\PBH, E_\elec) =
    \begin{dcases}
        \frac{2G^2M_\PBH^2E_\elec^2}{\hbar^2c^6} \,, & \frac{GM_\PBH E_\elec}{\hbar c^3} \leq 0.005 \\
        \Gamma^\text{num.}_{1/2}(M_\PBH, E_\elec)\,, & \frac{GM_\PBH E_\elec}{\hbar c^3} > 0.005 \,,
    \end{dcases}
\end{equation*}
and for photons this was
\begin{equation*}
    \Gamma_1^{}(M_\PBH, E_\phot) =
    \begin{dcases}
        \frac{64G^4M_\PBH^4E_\phot^4}{3\hbar^4c^{12}} \,, & \frac{GM_\PBH E_\phot}{\hbar c^3} \leq 0.005 \\
        \Gamma^\text{num.}_1(M_\PBH, E_\phot)\,, & \frac{GM_\PBH E_\phot}{\hbar c^3} > 0.005 \,,
    \end{dcases}
\end{equation*}
The cutoff value of \num{0.005} was chosen to avoid any large discontinuities at the function boundaries while still being consistent with the low-energy limit requirement.
The resulting PBH deposition efficiencies are given in Figure~\ref{fig:pbh-fdep}.

\subsection{Mass Distributions \label{subsec:method_mass-trunc}}

In practice, the integral in Eq.~\eqref{eq:F-frac} is performed numerically across \([\mu_\ten - 4\,\sigma_\ten,\mu_\ten + 4\,\sigma_\ten]\) using a simple trapezoidal integrator.
To compensate for distributions with varying widths, the integrator uses a constant 41 points across its range, giving accuracy comparable to a more thorough quadrature integrator.

The integration is performed within the larger range of \SIrange{e15}{e17}{\gram}, transformed to \numrange{-1}{1} in practice (as we normalise PBH masses by \SI{e16}{\gram} before taking the logarithm).
If either \(\mu_\ten - 4\,\sigma_\ten < -1\) or \(\mu_\ten + 4\,\sigma_\ten > 1\), the integration range is truncated to these bounds.
This is to prevent regions such as that below \SI{e15}{\gram} being included, which would break our assumption that PBH mass is effectively constant over its injection lifetime.
However, this truncation must be handled properly so PBH energy density is not unphysically lost.
At the low mass limit of \SI{e15}{\gram}, the lighter PBHs inject much more energy compared to heavier PBHs due to the \(M_\PBH^{-3}\) scaling as seen in Eq.~\eqref{eq:pbh-energy-inj}.
We compensate for truncation at this end by renormalising the distribution so that it retains its unity normalisation.
At the high-mass end the distribution is still truncated beyond \SI{e17}{\gram}, but it is not renormalised.
This is because these PBHs contribute negligibly to the total energy injection and so are safe to ignore.
If we were to renormalise the distribution in this sense, we would artificially inflate contributions from the low-mass end of the distribution, leading to stronger constraints.

This truncation correction is also used when evaluating log normal constraints through the extended monochromatic method detailed in Section~\ref{subsec:extending-constraints}.
This keeps the results consistent with the mass distribution implemented in the CMB analysis.

\subsection{Implementation}

To calculate the power spectra, the Boltzmann code \CLASS was used~\cite{Blas:2011rf}, which internally uses either \RECFAST~\cite{Seager:1999bc} or \HyRec~\cite{AliHaimoud:2010dx,AliHaimoud:2010ab} to solve for the ionisation history.
At the time of writing, \RECFAST was not written with non-standard DM physics in mind so \HyRec was used instead.
\HyRec moreover allows a more accurate incorporation of evaporating PBH physics into the recombination era.
Indeed, the fudge functions of \RECFAST are only tuned to (approximately) recover the standard recombination history, but need not be accurate when departing from it.
A beta version of \HyRec containing an implementation of WIMP DM physics was used to implement the PBH physics~\cite{Giesen:2012, AliHaimoud:2017}.
This was possible due to the similarities between evaporating PBHs and decaying WIMPs.
The PBH deposition efficiencies for each channel were converted to a bicubic spline and read in during runtime by \CLASS to be used by \HyRec.

The modified \CLASS was linked to the \Planck Likelihood Code~\cite{Aghanim:2015xee} to compute the corresponding \Planck likelihoods, which was then interfaced to \MultiNest~\cite{Feroz:2013hea}, a Monte Carlo implementation of the nested sampling algorithm~\cite{Feroz:2007kg,Feroz:2008xx}\footnote{For the interested reader, the resulting program has been made available at \url{https://github.com/hdp1213/pc_multinest}.}.
\MultiNest was chosen over more conventional MCMC samplers due to its robust convergence criterion.

\section{Results \label{sec:results}}

All results were produced using the `lite' high-\(\ell\) TT, TE and EE likelihood alongside the low-\(\ell\) TEB likelihood.
The choice of these likelihoods is in keeping with the baseline \Planck analysis.
The `lite' high-\(\ell\) likelihood has many \Planck nuisance parameters marginalised out, and was used in place of the full likelihood in the interests of computation time.
The one nuisance parameter left in this likelihood, \(y_\text{cal}\), is an overall normalisation of the \Planck data which we hold fixed throughout this analysis.

We also use a \LCDM parametrisation similar to \Planck, choosing (\(\Omegab h^2\), \(\Omegac h^2\), \(100\theta_\text{s}\), \(\tau\), \(\ln(10^{10}A_\text{s})\), \(\ns\)) as the set of base parameters.
We note that our choice of \(100\theta_\text{s}\) differs from the \Planck \(100\theta_\text{MC}\) as we do not use CosmoMC in our analysis and this is a parameter exclusive to that program.
Table~\ref{tab:lcdm-parameters} contains the scan ranges and fixed values used for these \LCDM parameters, alongside \(y_\text{cal}\) and the PBH parameters.
The fixed \LCDM values are our own best-fit values using a vanilla \LCDM model, and only differ from the \Planck values at most by less than 0.2 sigma.
The reason we use our own values instead of the \Planck ones was to have certainty that the best-fit values used corresponded to the maximum of the likelihood when using a vanilla \LCDM model.
Although we do reproduce \Planck posterior distributions in the vanilla model, slight differences in Boltzmann codes used led to these slight discrepancies which are worth accounting for especially when incorporating new physics.

\begin{table}[b]
  \centering
  \caption{Priors and fixed values used for \LCDM, the \Planck calibration parameter \(y_\text{cal}\) and PBH parameters throughout this work.}
  \label{tab:lcdm-parameters}
  \begin{tabular}{>{\(}c<{\)} *{2}{c}}
    \toprule
    \text{Parameter} & Uniform prior & Fixed value \\
    \midrule
    \Omegab h^2 & \([0.016, 0.028]\) & \num{0.022220} \\
    \Omegac h^2 & \([0.108, 0.130]\) & \num{0.11983} \\
    100\theta_\text{s} & \([1.039, 1.043]\) & \num{1.041788} \\
    \tau & \([0.01, 0.15]\) & \num{0.0803} \\
    \ln(10^{10}A_\text{s}) & \([2.98, 3.20]\) & \num{3.0957} \\
    \ns & \([0.92, 1.04]\) & \num{0.96352} \\
    \midrule
    y_\text{cal} & --- & \num{1.00047} \\
    \midrule
    \log_\ten(f_\PBH) & \([-8, 0]\) & --- \\
    \mu_\ten & \([-1, 1]\) & --- \\
    \log_\ten(\sigma_\ten) & \([-2, 1]\) & --- \\
    \bottomrule
  \end{tabular}
\end{table}

The exclusion regions presented in this section are the inverse of the Bayesian \SI{95}{\percent} credible regions of the 2D marginalised posteriors of each pair of parameters plotted.
This means that for a \SI{95}{\percent} exclusion region, the integrated probability of parameters lying in such regions is only \SI{5}{\percent}.

\subsection{Delta Mass Distribution}

\subsubsection{Recombination Effects}

\begin{figure}[t!]
  \centering
  \begin{subfigure}[t]{0.49\textwidth}
    \centering
    \includegraphics[height=2.2in]{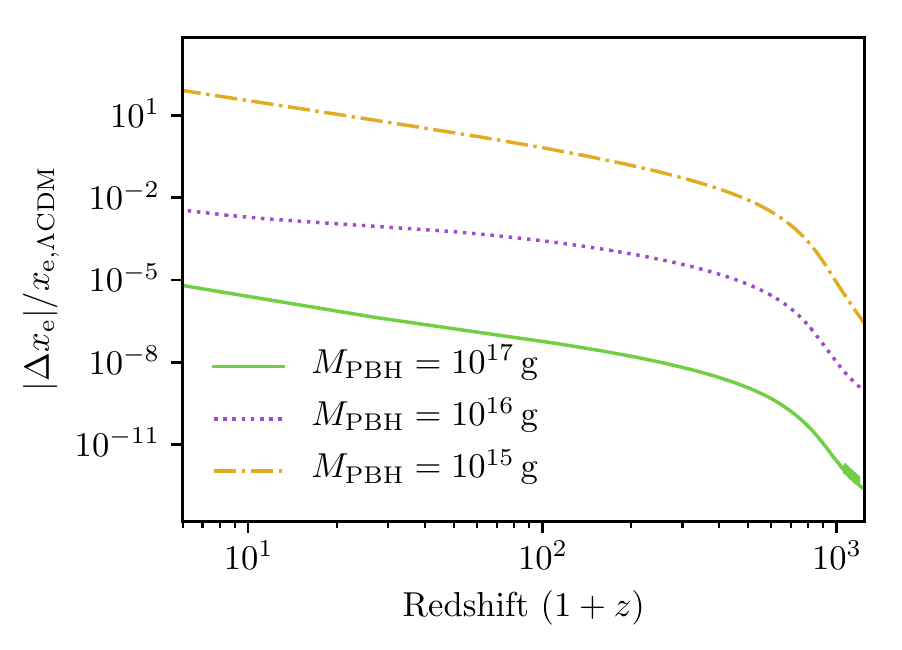}
    \label{subfig:pbh-xe}
  \end{subfigure}%
  ~
  \begin{subfigure}[t]{0.49\textwidth}
    \centering
    \includegraphics[height=2.2in]{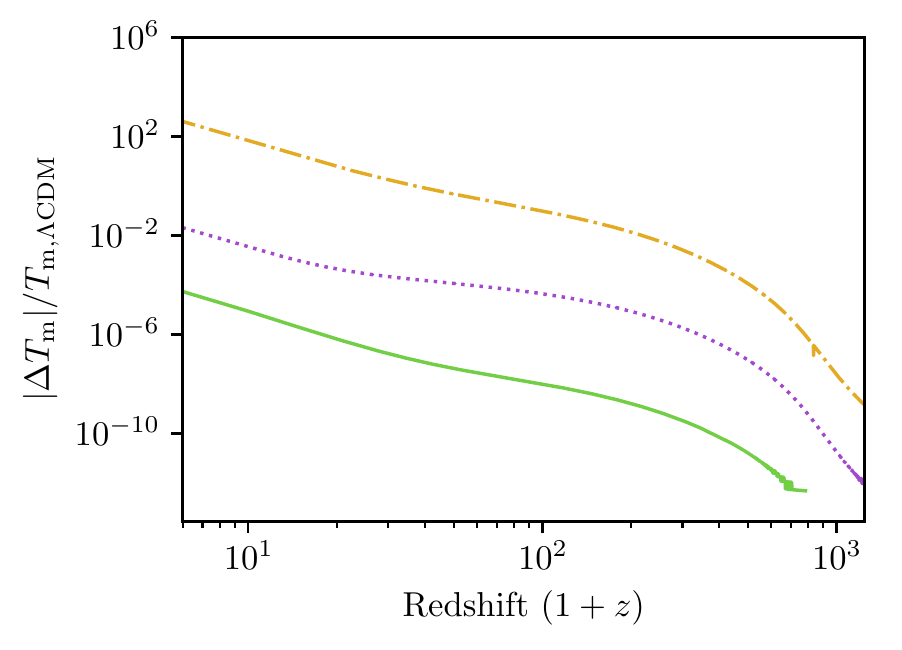}
    \label{subfig:pbh-Tm}
  \end{subfigure}
  \vspace{-2.0em}
  \caption{Relative differences between ionisation \(x_\elec\) and matter temperature \(T_\mat\) histories for evaporating PBHs compared to a vanilla model with no PBHs. Here, reionisation is not shown. \(f_\PBH\) is set to \num{e-7} across all curves. It can be seen that the effect of evaporating PBHs roughly scales as \(M_\PBH^{-3}\).}
  \label{fig:pbh-histories}
\end{figure}

Adding evaporating PBHs into the early Universe changes the standard ionisation and temperature histories.
Disregarding reionisation, Figure~\ref{fig:pbh-histories} shows how the free electron fraction \(x_\elec\) and matter temperature \(T_\mat\) change with differing PBH masses.
For each plot, the fraction \(f_\PBH\) is set to \num{e-7}.
We see that the absolute value of the differences roughly scales as \(M_\PBH^{-3}\), \emph{i.e.} that each history decreases by three orders of magnitude for each PBH mass increase of one.
We also see the largest difference occurs at later redshifts.
In the standard \LCDM model without reionisation, no extra ionisation or heating happens after recombination.
With the addition of PBHs, both ionisation and heating increase relative to the vanilla \LCDM model.

\begin{figure}[t!]
  \centering
  \begin{subfigure}[t]{0.49\textwidth}
    \centering
    \includegraphics[height=2.2in]{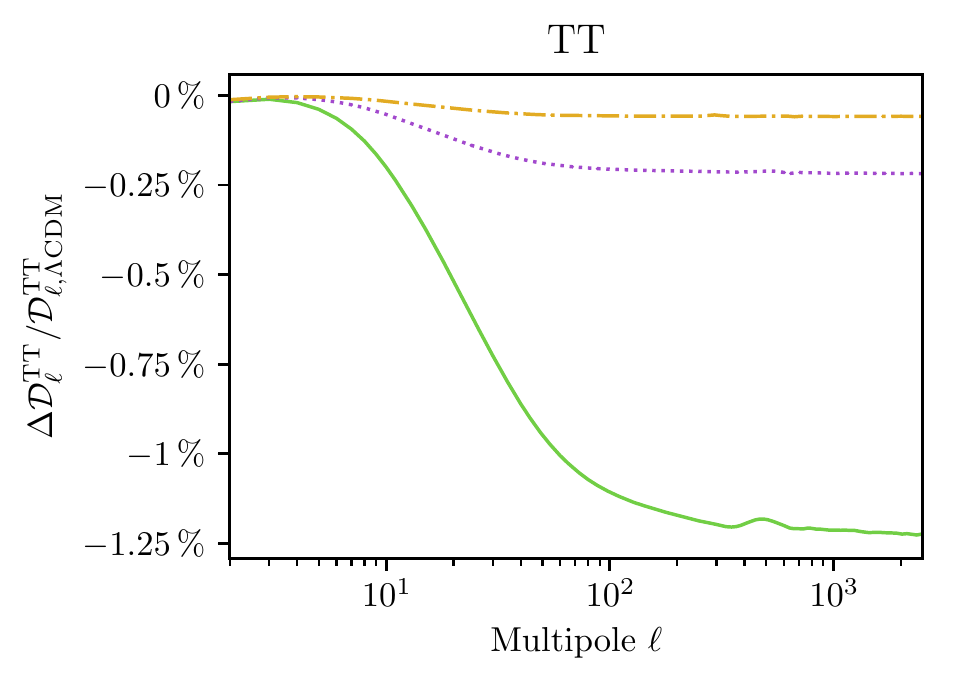}
    \label{subfig:pbh-TT}
  \end{subfigure}%
  ~
  \begin{subfigure}[t]{0.49\textwidth}
    \centering
    \includegraphics[height=2.2in]{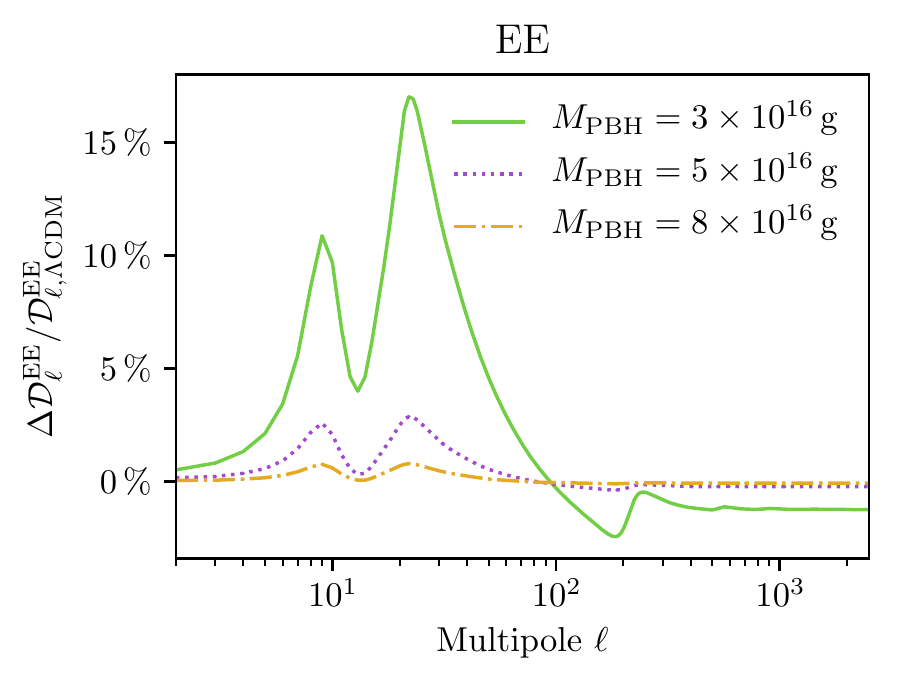}
    \label{subfig:pbh-EE}
  \end{subfigure}
  \vspace{-2.0em}
  \caption{Fractional change in CMB temperature and \(E\)-mode polarisation power spectra under the addition of evaporating PBHs. Here, PBH dark matter is assumed to make up \SI{10}{\percent} of dark matter, \emph{i.e.}, \(f_\PBH = 0.1\) for all masses considered. Here again, the effect scales roughly as \(M_\PBH^{-3}\).}
  \label{fig:pbh-spectra}
\end{figure}

Similarly, Figure~\ref{fig:pbh-spectra} shows the effect that various sizes of evaporating PBHs have on the temperature and \(E\)-mode polarisation power spectra if they were to constitute \SI{10}{\percent} of the dark matter energy density.
For both spectra, the small scale fluctuations are suppressed by the addition of PBHs.
In the polarisation spectrum, we see that mid-range fluctuations (\(\ell \sim 20\)) are increased by a large amount relative to the rest of the spectrum.
However, for \(M_\PBH = \SI{3e16}{\gram}\) a \SI{10}{\percent} fraction of PBHs would be completely excluded by accurate measurements of the EE power spectrum.

All plots in this section have been created with \LCDM parameters set to the fixed values given in Tab.~\ref{tab:lcdm-parameters}.

\subsubsection{Parameter Degeneracies}

As well as examining the effects PBHs have on the ionisation and thermal histories, we can consider how they affect the base \LCDM parameters.
For this, we fix \(M_\PBH\) to \SI{e16}{\gram} and allow the PBH fraction \(f_\PBH\) to vary.
For our choice of log-prior, the \LCDM parameters are virtually unaffected, since the bulk of the volume of the posterior falls into a region where the PBHs are essentially invisible to the data.
As such, we instead use a flat prior of \([0,1]\) for this section.
By only allowing one PBH parameter to vary, we also remove any potential degeneracies multiple parameters may have with each other: here, \(f_\PBH\) is the sole parameter responsible for increasing the energy injection from PBHs.

\begin{figure}[t!]
  \centering
  \includegraphics[width=6.1in]{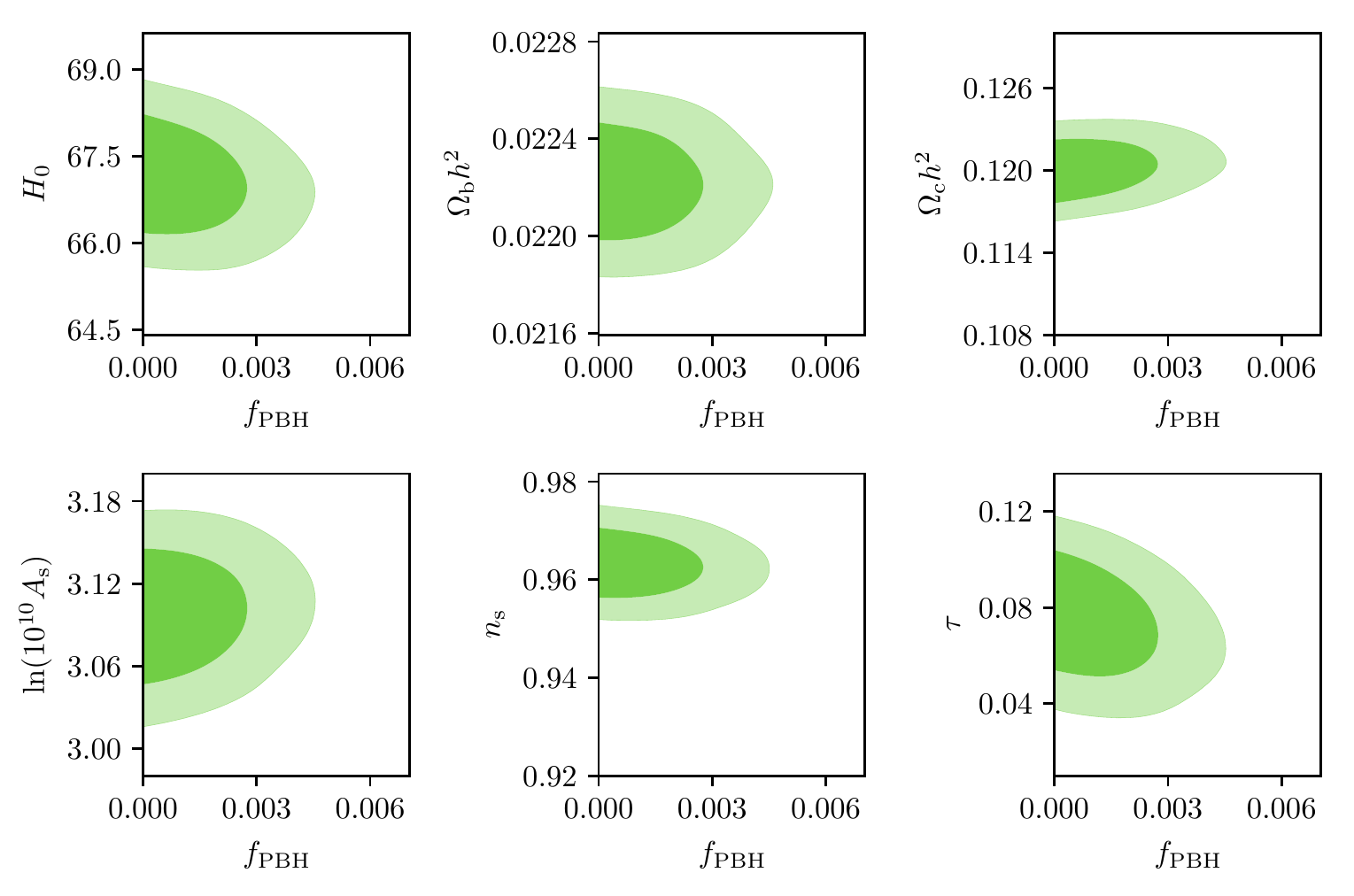}
  \caption{\SI{68}{\percent} (light) and \SI{95}{\percent} (dark) Bayesian credible regions for \(f_\PBH\) versus each \LCDM parameter. PBH mass was fixed to \(M_\PBH = \SI{e16}{\gram}\).}
  \label{fig:pbh-degen}
\end{figure}

\begin{figure}[t!]
  \centering
  \begin{subfigure}[t]{0.49\textwidth}
    \centering
    \includegraphics[width=2.6in]{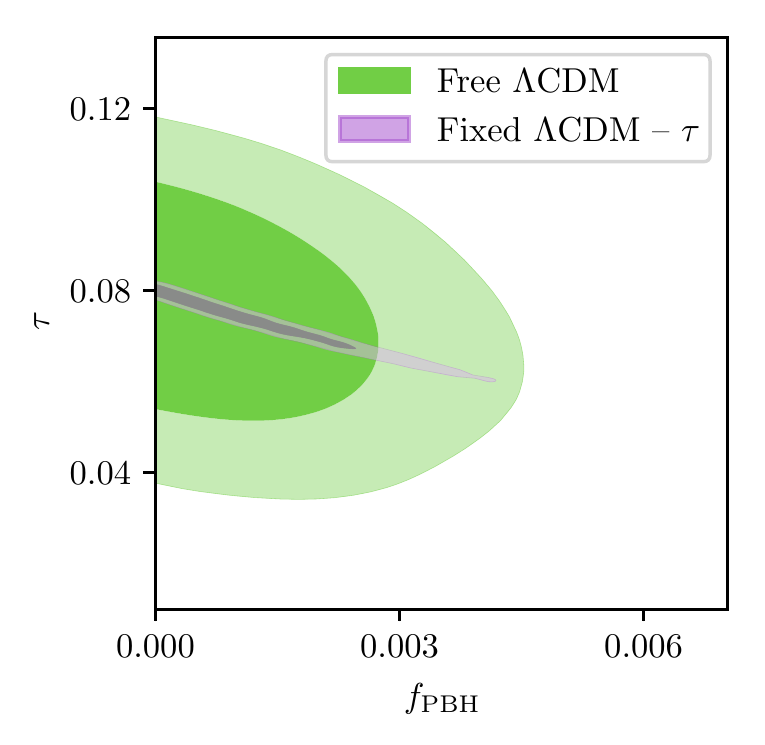}
    \label{subfig:pbh-tau-degen}
  \end{subfigure}%
  ~
  \begin{subfigure}[t]{0.49\textwidth}
    \centering
    \includegraphics[width=2.6in]{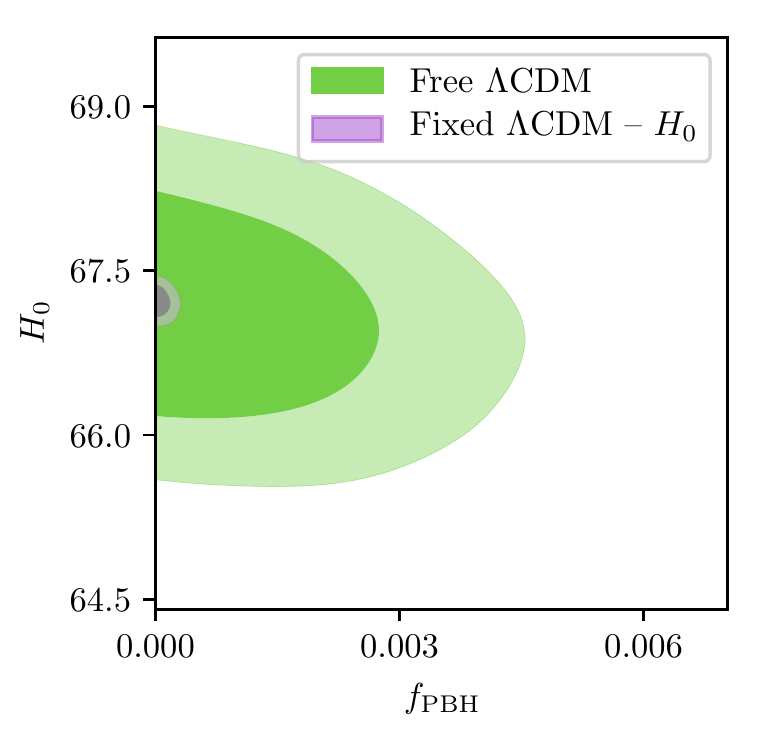}
    \label{subfig:pbh-H0-degen}
  \end{subfigure}
  \vspace{-2.0em}
  \caption{Comparison between \SI{68}{\percent} (light) and \SI{95}{\percent} (dark) Bayesian credible regions for \(f_\PBH\) versus \(\tau\) (left) and \(f_\PBH\) vs \(H_0\) (right) with free and fixed \LCDM parameters. PBH mass was fixed to \(M_\PBH = \SI{e16}{\gram}\).}
  \label{fig:pbh-param-degen}
\end{figure}

Figure~\ref{fig:pbh-degen} shows the 2D marginalised posteriors for \(f_\PBH\) against each \LCDM parameter.
The parameters that show the most degeneracy with \(f_\PBH\) are \(\tau\) and \(H_0\), with weaker dependencies exhibited by \(\ln(10^{10} A_\text{s})\) and \(\Omegac h^2\).
We have already seen that the presence of decaying PBHs enhances the ionisation fraction at low redshifts.
The same effect is also achieved by increasing the reionisation optical depth \(\tau\), so as the two parameters effectively cancel each other's contributions, we expect them to be anticorrelated.
This relationship can be more clearly examined if we fix all other \LCDM parameters and only allow \(\tau\) to vary alongside \(f_\PBH\).
Figure~\ref{fig:pbh-param-degen} shows the resulting 2D marginalised posterior when compared to the original in Fig.~\ref{fig:pbh-degen}.
We see that when all other \LCDM parameters are fixed, the extent of the posterior along the \(f_\PBH\) axis is almost that of the posterior with freed \LCDM parameters.

Similarly, we see the value of \(H_0\) decreases slightly as \(f_\PBH\) increases.
We note that the value of \(H_0\) inferred from CMB data in the \(\Lambda\)CDM model is already in tension with local astrophysical measurements~\cite{Riess:2016jrr,Aghanim:2018eyx}, and that the inclusion of PBHs does not help to reconcile this difference.
This slight degeneracy is somewhat accidental: the effect that changing \(f_\PBH\) has on the CMB power spectrum is similar enough to those made by changing \(H_0\) such that the addition of \(f_\PBH\) affects the inference of \(H_0\).
As evidence of this, Fig.~\ref{fig:pbh-param-degen} shows how \(H_0\) otherwise remains constant when all other \LCDM parameters are fixed.
The sloping seen in the free \LCDM posterior can then only be due to a slight correlation between \(H_0\) and \(\tau\): as the optical depth decreases, then so does the apparent Hubble constant.

\subsubsection{Lambda-CDM Limits}

\begin{figure}[!t]
  \centering
  \includegraphics[height=4.2in]{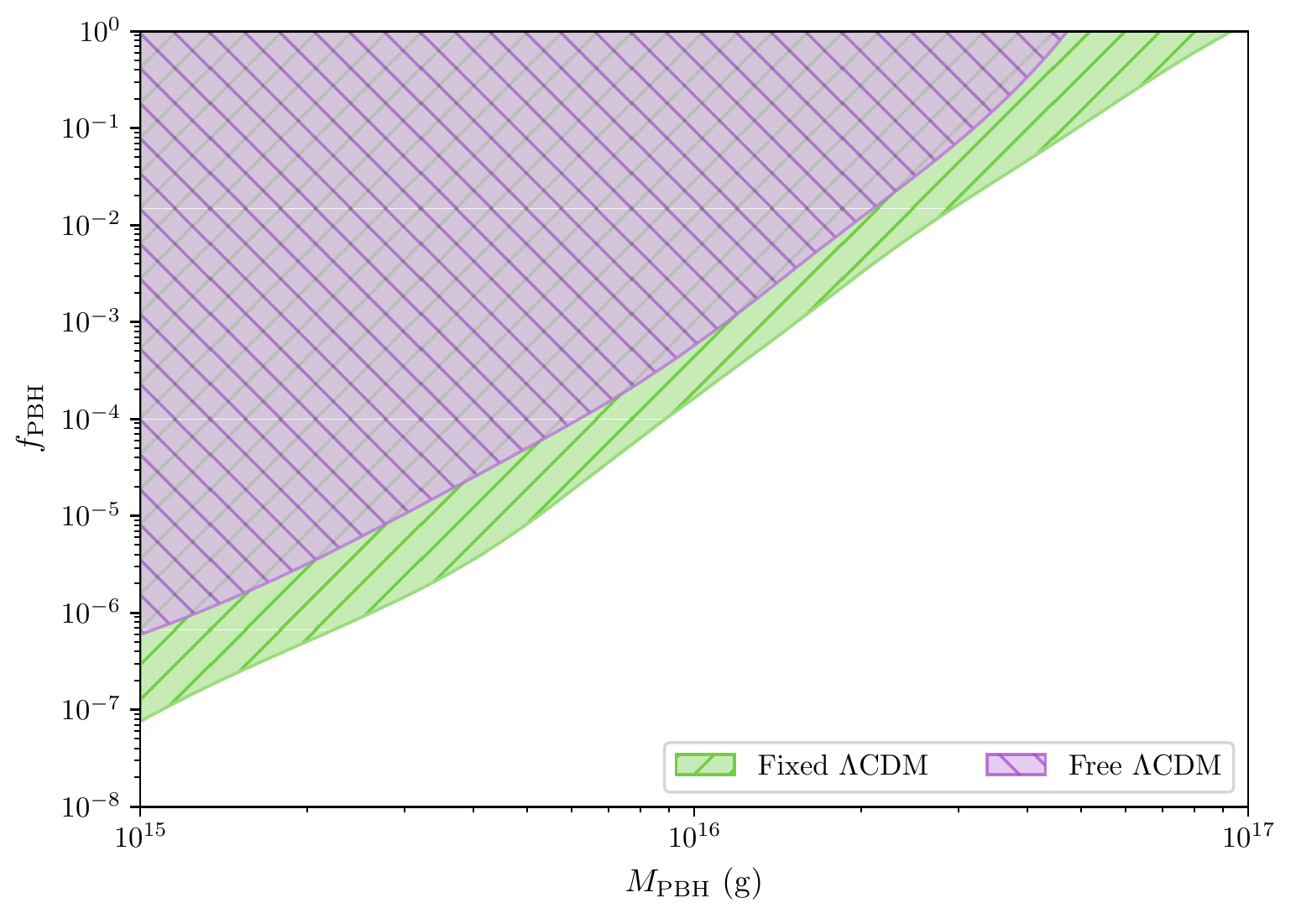}
  \caption{\SI{95}{\percent} exclusion regions for both free and fixed \LCDM parameters. Both bounds apply to a monochromatic mass distribution.}
  \label{fig:clark-limit}
\end{figure}

We now move on to consider the effect allowing base \LCDM parameters to vary has on bounds on \(f_\PBH\).
Here, we improve upon the analysis presented in~\citetalias{Clark:2016nst} by relaxing the assumption that the variation of \LCDM parameters is negligible enough to ignore.
Figure~\ref{fig:clark-limit} shows how allowing \LCDM parameters to vary lessens the exclusion bounds by about an order of magnitude across the mass range considered.
From our previous analysis, we can say this is predominantly due to the relaxation of the optical depth parameter \(\tau\).
As a smaller optical depth to reionisation counteracts the effects of an increase in early universe ionisation, it makes sense then that allowing \(\tau\) to vary from its currently-measured value allows larger energy injections and thus weakens the constraints placed by CMB measurements.

More interestingly, the exclusion regions relax more for either end of the mass spectrum compared to the middle region to a point where PBHs with mass \(\gtrsim\SI{5e16}{\gram}\) can feasibly account for all of the dark matter.
At this end, PBHs are too heavy to inject meaningful amounts of radiation, and so the base \LCDM parameters are able to vary to fully compensate for this.
The fact that the relaxation of the constraint is less pronounced around \(M_\PBH \sim \SI{e16}{\gram}\) may be due to a smaller degeneracy with standard \LCDM parameters, due to a different redshift dependence of the ionization fraction.

As a consistency check, we also find that the fixed \LCDM exclusion region agrees well with the \SI{95}{\percent} exclusion region presented in~\citetalias{Clark:2016nst}.

\subsection{Extended Mass Distributions}

We will now examine the effect that an extended mass distribution of PBHs has on recombination.
We consider constraints placed on the width \(\sigma_\ten\) of a log-normal distribution, as well as constraints on \(f_\PBH\) for a variety of different mass distributions.
Throughout this section, we have implicitly made the conversion from the distribution mean \(\mu_\ten\) appearing in Eq.~\eqref{eq:log-norm-dist} to a physical mass \(M_\PBH\) which we will refer to as the mean mass of the distribution.

\subsubsection{Constraints on Distribution Width}

\begin{figure}[t]
  \centering
  \includegraphics[height=3.8in]{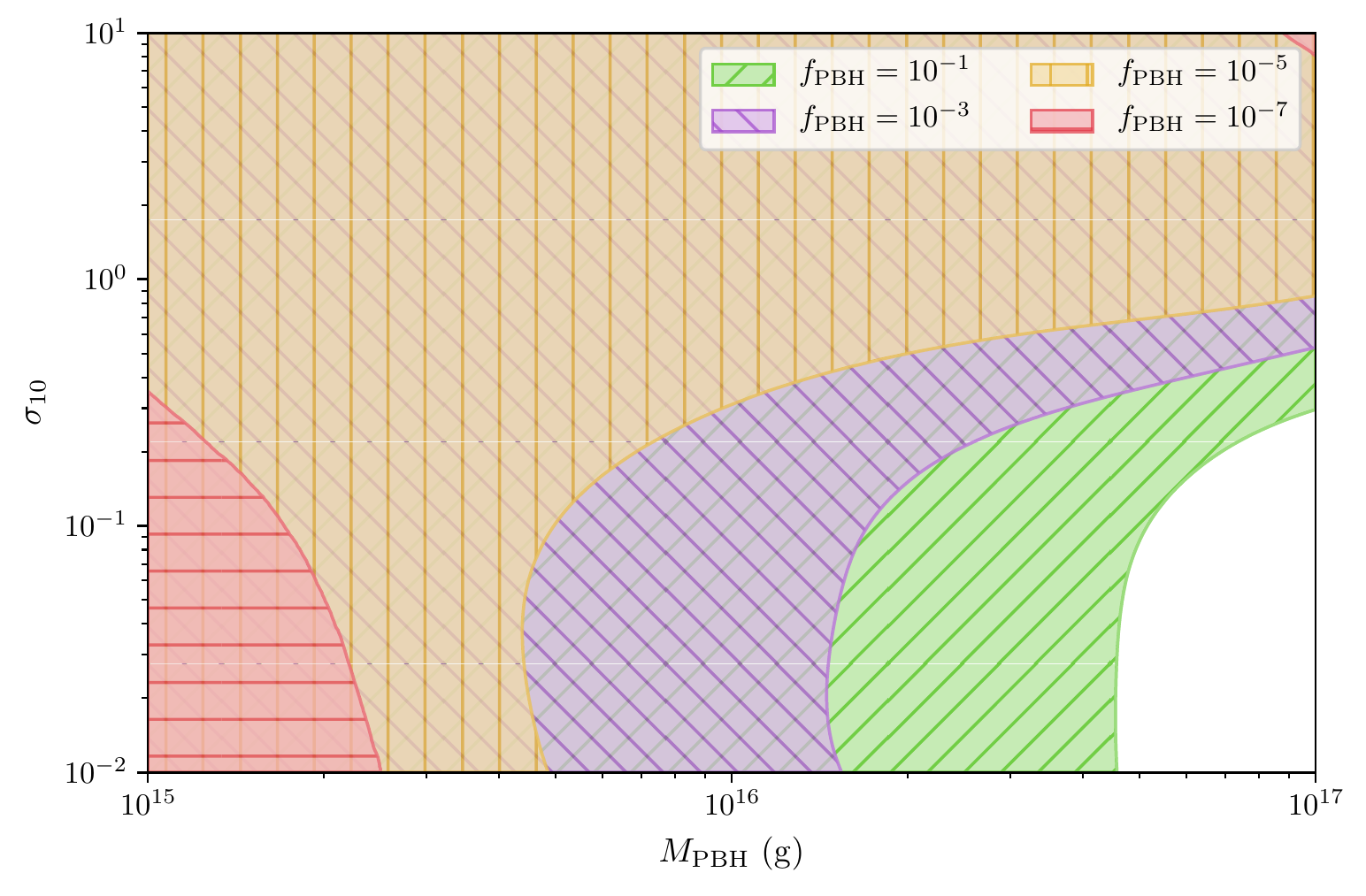}
  \caption{\SI{95}{\percent} exclusion regions for log-normal PBH mass distributions with fixed \(f_\PBH\) and \LCDM parameters.
  All regions overlap and except for \(f_\PBH = \num{e-7}\) extend to the upper left corner of the plot.}
  \label{fig:frac-limit}
\end{figure}

\begin{figure}[t]
  \centering
  \includegraphics[height=3.8in]{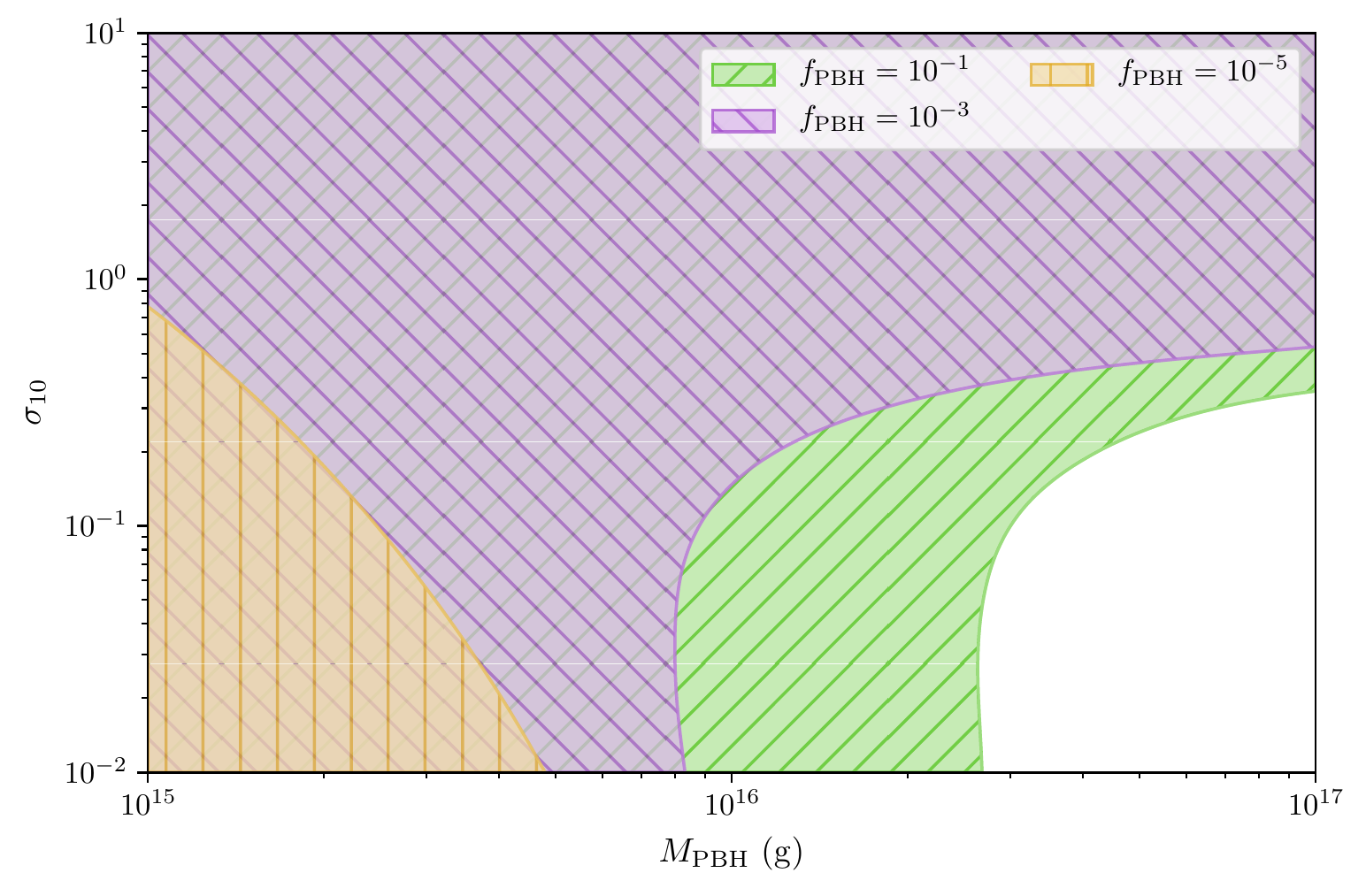}
  \caption{\SI{95}{\percent} exclusion regions for log-normal PBH mass distributions with fixed \(f_\PBH\) and free \LCDM parameters. All regions overlap and except for \(f_\PBH = \num{e-5}\) extend to the upper left corner of the plot}
  \label{fig:frac-limit-lcdm}
\end{figure}

First, we consider the family of log-normal distributions spanning a delta function-like mass distribution at \(\sigma_\ten = \num{e-2}\) to an effectively uniform distribution at \(\sigma_\ten = 10\).
Figure~\ref{fig:frac-limit} gives the \SI{95}{\percent} exclusion regions for a variety of log-normal distributions for different values of \(f_\PBH\), with all \LCDM parameters fixed to their values in Tab.~\ref{tab:lcdm-parameters}.
Similarly, Figure~\ref{fig:frac-limit-lcdm} shows the \SI{95}{\percent} exclusion regions for the case where all \LCDM parameters are allowed to vary.

In Fig.~\ref{fig:frac-limit}, the exclusion regions for \(f_\PBH = \num{e-1}\), \num{e-3} and \num{e-5} completely exclude the region defined by \(\sigma_\ten \geq 1\).
At such high values of \(\sigma_\ten\), the distribution spans more orders of magnitude than present in the mass range \SIrange{e15}{e17}{\gram}, leading to a uniform distribution that behaves independently of its mean mass value.
However, we see for \(f_\PBH = \num{e-7}\) that distributions with \(\sigma_\ten \geq 1\) are allowed.
As this region does not depend on \(M_\PBH\), it can either be completely excluded or allowed for some selection of \(f_\PBH\).
This then leads to the idea of a ``critical value'' of \(f_\PBH\) at which this transition occurs, which will be discussed later on.

By examining the way that the non-shaded permitted regions for \(f_\PBH = \num{e-1}\), \num{e-3} and \num{e-5} change shape, we see relationships between \(f_\PBH\) and the distribution shape parameters.
Namely, we see that lowering the PBH fraction predominantly allows lighter mean PBH masses, rather than allowing a wider distribution.
Another way of saying this is that the distribution's width is more tightly constrained by \(f_\PBH\) compared to its mean.
This holds true up to the critical value of \(f_\PBH\), where the exclusion region changes shape drastically to allow the \(\sigma_\ten \geq 1\) region.

Examining Fig.~\ref{fig:frac-limit-lcdm}, we see that allowing the base \LCDM parameters to vary relaxes the exclusion bounds, as one would expect.
Similarly to the fixed \LCDM case, we see that when reducing \(f_\PBH\), the largest relaxation of the bounds is along the mean PBH mass axis.
We also see that the \(f_\PBH = \num{e-5}\) region turns from a closed bound for fixed \LCDM to an open bound when these parameters are freed, implying that the critical \(f_\PBH\) value for uniform distributions increases when \LCDM parameters are allowed to vary.
We will examine this in more detail later.
No constraint is shown for \(f_\PBH = \num{e-7}\), as it is completely unconstrained when \LCDM parameters are allowed to vary.

For the values of \(f_\PBH\) that exclude \(\sigma_\ten \geq 1\), it is apparent across both figures that these regions have the same characteristic shape to them.
In particular, the exclusion bounds appear to reach a limiting \(\sigma_\ten\) asymptotically as \(M_\PBH\) increases.
This may suggest that each PBH fraction has some maximum value of \(\sigma_\ten\) they constrain in the limit of high PBH mass.
However, this limit is not asymptotic as the effective constraining power comes solely from the low-mass end, with larger PBH masses inject ever-lower amounts of energy.
This suggests that the constraint will slowly become weaker (constraining less of the parameter space) if it were to be extended to larger masses.

Similarly, the exclusion bounds appear to reach a limiting value of \(M_\PBH\) as \(\sigma_\ten\) decreases.
In contrast, this behaviour is asymptotic in that the smaller \(\sigma_\ten\) gets, the minimally allowed mean PBH mass \(M_\PBH\) will remain constant.
This is because as \(\sigma_\ten\) decreases, the log-normal distribution approaches that of a monochromatic one.
As such, the constraint will approach the monochromatic constraint that we have already seen in Fig.~\ref{fig:clark-limit}.

\subsubsection{Constraints on PBH Fraction}

We may also look at the constraints on \(f_\PBH\) that a selection of distribution widths give us.
For this we consider a uniform distribution, a log-normal distribution with fixed width, a delta distribution and a log-normal distribution with variable width which is then integrated over, weighted by its posterior probability.
A width of \(\sigma_\ten = 0.3\) was chosen for the log-normal distribution as this corresponds to the region in Fig.~\ref{fig:frac-limit} where the most amount of variance was seen across PBH mass.
Note that we approximate the uniform distribution by a wide log-normal distribution;
as the mass range under consideration only spans two orders of magnitude, setting \(\sigma_\ten = 10\) corresponds to an effectively uniform mass distribution.
Distribution cut-off by the low-mass limit \(M = \SI{e15}{\gram}\) is handled as described in Section~\ref{subsec:method_mass-trunc}.

\begin{figure}[t]
  \centering
  \includegraphics[height=3.5in]{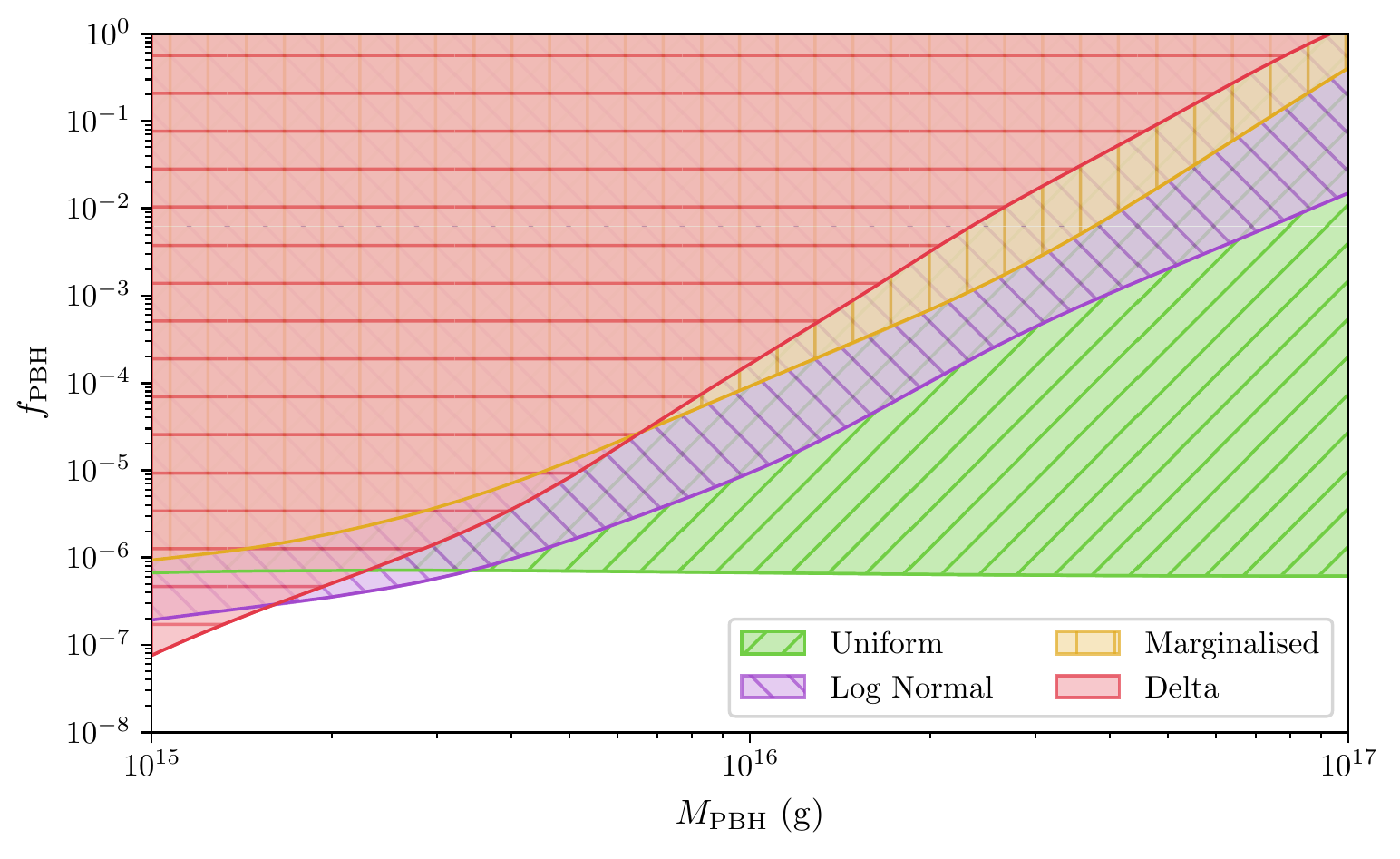}
  \caption{\SI{95}{\percent} exclusion regions for varying PBH mass distributions with \LCDM parameters fixed. The uniform distribution is a log-normal distribution with \(\sigma_\ten = 10\), the log-normal distribution has \(\sigma_\ten = 0.3\), and the marginalised distribution is a log-normal marginalised over \(\sigma_\ten\).}
  \label{fig:dist-limit}
\end{figure}

\begin{figure}[t]
  \centering
  \includegraphics[height=3.5in]{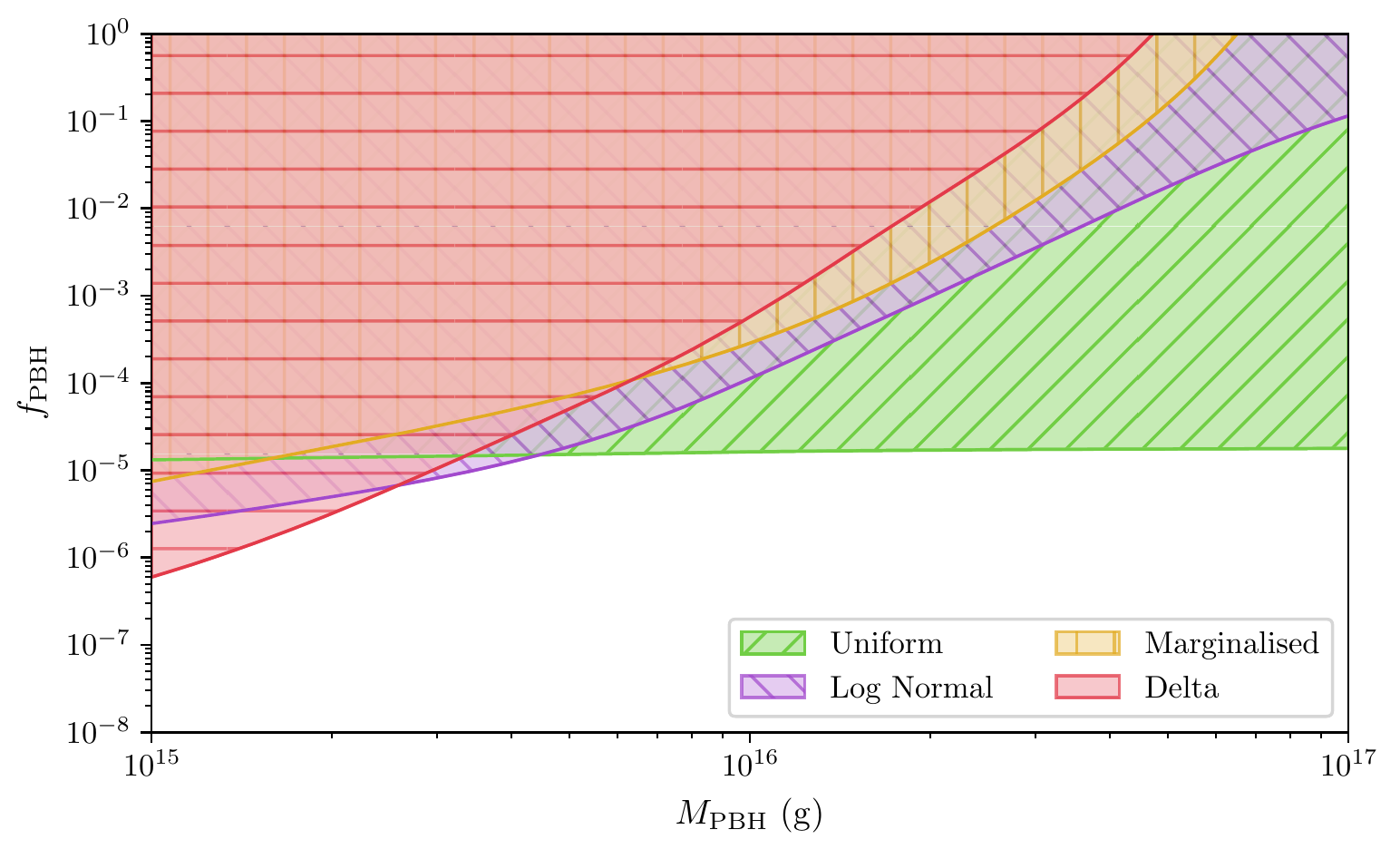}
  \caption{\SI{95}{\percent} exclusion regions for varying PBH mass distributions with \LCDM parameters freed. The mass distributions used are the same as those in Figure~\ref{fig:dist-limit}.}
  \label{fig:dist-limit-lcdm}
\end{figure}

Figure~\ref{fig:dist-limit} shows \SI{95}{\percent} bounds for three different mass distributions and a marginalised log-normal distribution, where \LCDM parameters are fixed to their best-fit given in Tab.~\ref{tab:lcdm-parameters}, whilst Figure~\ref{fig:dist-limit-lcdm} allows them to vary.

What was previously alluded to during the last section without being properly explained was the concept of a critical value of \(f_\PBH\) for which the exclusion bounds drastically change shape between fully excluding to fully allowing the \(\sigma_\ten \geq 1\) region.
By changing the parameter space to fix \(\sigma_\ten\) while allowing \(f_\PBH\) to vary, we are able to see this from a different perspective: that the critical \(f_\PBH\) is simply the value at which a uniform distribution becomes excluded.
This can be seen in the constraints of Figs.~\ref{fig:dist-limit} and \ref{fig:dist-limit-lcdm}, with a value of \(f_\PBH < \num{6.7e-7}\) for fixed \LCDM and \(f_\PBH < \num{1.6e-5}\) for free \LCDM.
These values are consistent with what we saw in the last section where the ``critical value'' lay between \numrange{e-7}{e-5} for fixed \LCDM and \numrange{e-5}{e-3} for free \LCDM.
As well as this, we see the choice of \(\sigma_\ten = 10\) for the log-normal is sufficient to treat it as a uniform mass distribution, as its bounds are independent of the PBH mass across both figures.

The remaining mass distributions follow the more characteristic triangular exclusion region seen previously with a delta mass distribution.
Compared to a delta mass distribution, both the marginalised and log-normal distributions have tighter exclusion bounds for heavier mean PBH mass.
As these distributions have finite widths, both incorporate lighter PBHs which inject more energy proportional to \(M_\PBH^{-3}\).
As a consequence, contributions from the medium to high PBH mass range are effectively dominated by PBHs with masses roughly 0.3 orders of magnitude lower for the log-normal distribution.
The marginalised distribution imposes weaker limits as by marginalising over the range of widths \([\num{e-2},10]\) in log space, this distribution incorporates thinner, more delta-like distributions.

Conversely, at low PBH mass both the log-normal and marginalised distributions give weaker bounds than the delta distribution.
Compared to a delta distribution, a log-normal distribution has much less probability centred around its mean value, meaning there is less contribution given to PBHs of the same mass across these distributions.
In a similar manner to the previous discussion, the inclusion of heavier PBHs in the non-truncated tail of log-normal distributions contributes negligibly to the total energy injection due to the inverse cube dependence on their mass.
Consequently, the total energy injection will be less than that of a pure delta distribution which is why we see this overlapping behaviour at the low-mass end of the spectrum.

Comparing both the fixed and free \LCDM plots, we see again that allowing the base \LCDM parameters to vary relaxes the exclusion bounds across all mass distributions.
The reduction is roughly an order of magnitude across the PBH mass range, just like for the previously-examined delta mass distribution.

The net effect of spreading out the PBH mass distribution from a delta distribution in this way is to increase the exclusion bounds for masses \(M_\PBH \gtrsim \SI{3e15}{\gram}\) by a few orders of magnitude and decrease exclusion bounds for lighter masses by considerably less.

\subsubsection{Validation of Extending Monochromatic Mass Constraints}

Before considering extending other monochromatic mass constraints to compare them with the constraints presented in this paper, it is worth examining how well this technique performs by comparing the extension of our monochromatic mass results in Fig.~\ref{fig:clark-limit} to our own computed constraints.

\begin{figure}[t!]
  \centering
  \begin{subfigure}[t]{0.49\textwidth}
    \centering
    \includegraphics[height=2.4in]{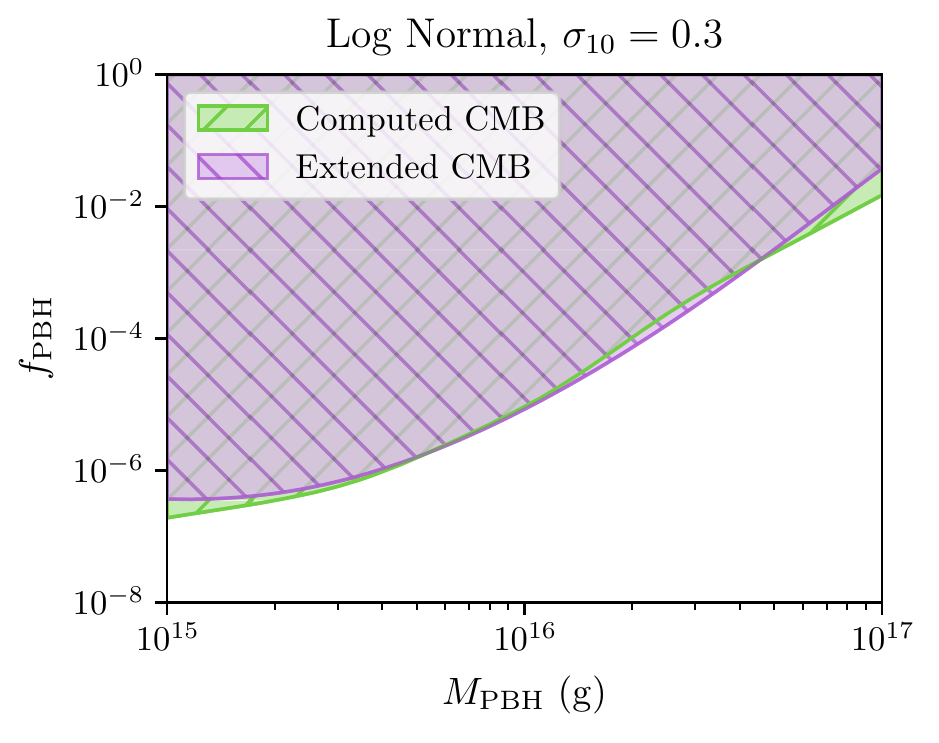}
    \label{subfig:cmb-self-lognorm}
    \vspace{-2em}
  \end{subfigure}
  \begin{subfigure}[t]{0.49\textwidth}
    \centering
    \includegraphics[height=2.4in]{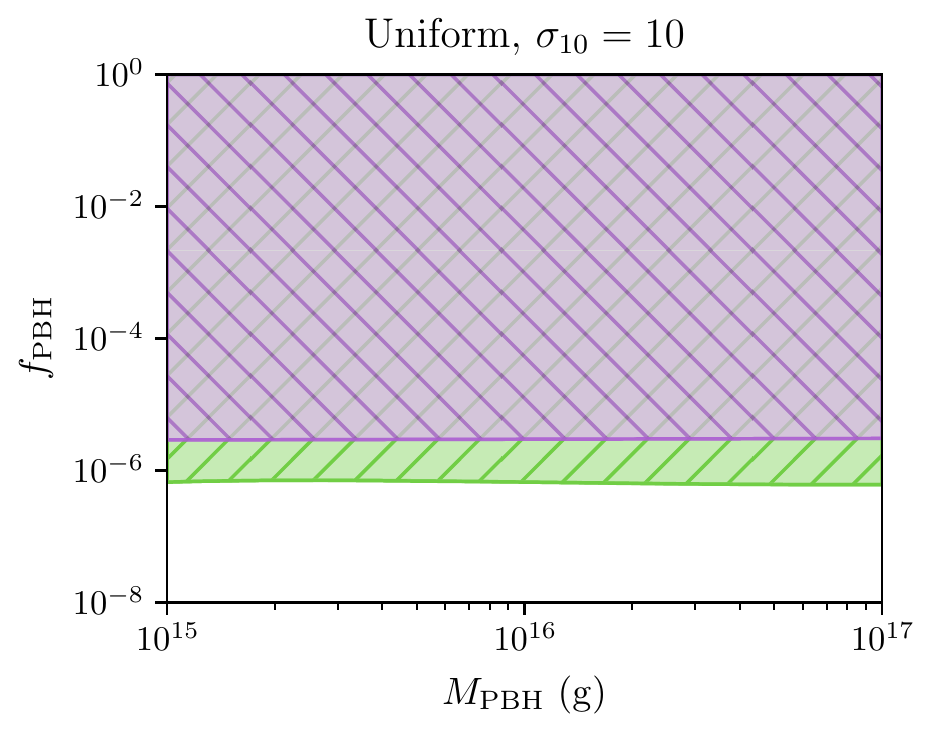}
    \label{subfig:cmb-self-uniform}
  \vspace{-2.0em}
  \end{subfigure}
  \caption{Comparison of \SI{95}{\percent} exclusion regions derived from the computation outlined in this paper, and by extending the monochromatic mass constraints in Fig.~\ref{fig:clark-limit} as outlined in Ref.~\cite{Carr:2017jsz}. Here, both results are derived with the base \LCDM parameters fixed to the values given in Table~\ref{tab:lcdm-parameters}. The monochromatic mass distribution is not shown, as the extended constraints would remain unchanged.}
  \label{fig:cmb-self-comp}
\end{figure}

We follow the outline given in Sec.~\ref{subsec:extending-constraints} to extend the fixed \LCDM constraint given in Fig.~\ref{fig:clark-limit} by using it as the \(f_\text{max}(M_\PBH)\).
Figure~\ref{fig:cmb-self-comp} shows the comparison of the two methods for the fixed \LCDM parameter constraints.
Due to the way that the limits are extended, we recover the original monochromatic constraint as \(\sigma_\ten\to0\), hence why we do not show this comparison.
We see that extending our monochromatic constraints to a log-normal with \(\sigma_\ten = 0.3\) gives excellent agreement with the fully-computed limits.
Interestingly, there is a slight disagreement between the computed and extended constraints for a uniform mass distribution, with the extended result giving a maximum allowed value of \(f_\PBH\) roughly \(\sim 4.5\) times larger than the computed bounds.

\subsubsection{Comparisons with Gamma Ray Background Measurements}

\begin{figure}[t!]
  \centering
  \begin{subfigure}[t]{0.49\textwidth}
    \centering
    \includegraphics[height=2.4in]{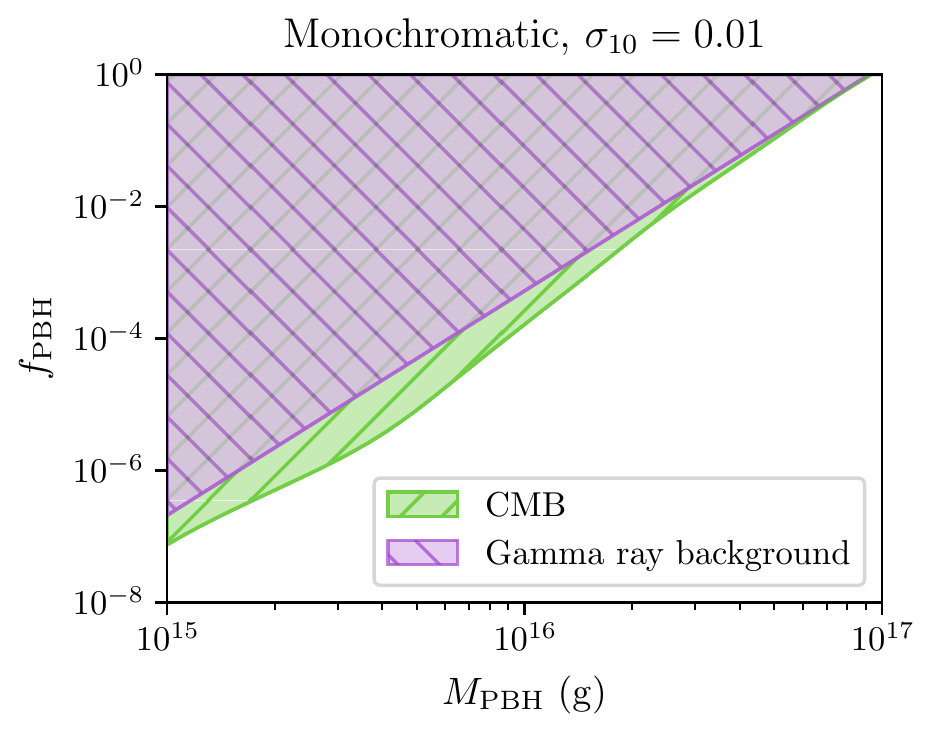}
    \label{subfig:gamma-cmb-delta}
    \vspace{-2em}
  \end{subfigure}%
  ~
  \begin{subfigure}[t]{0.49\textwidth}
    \centering
    \includegraphics[height=2.4in]{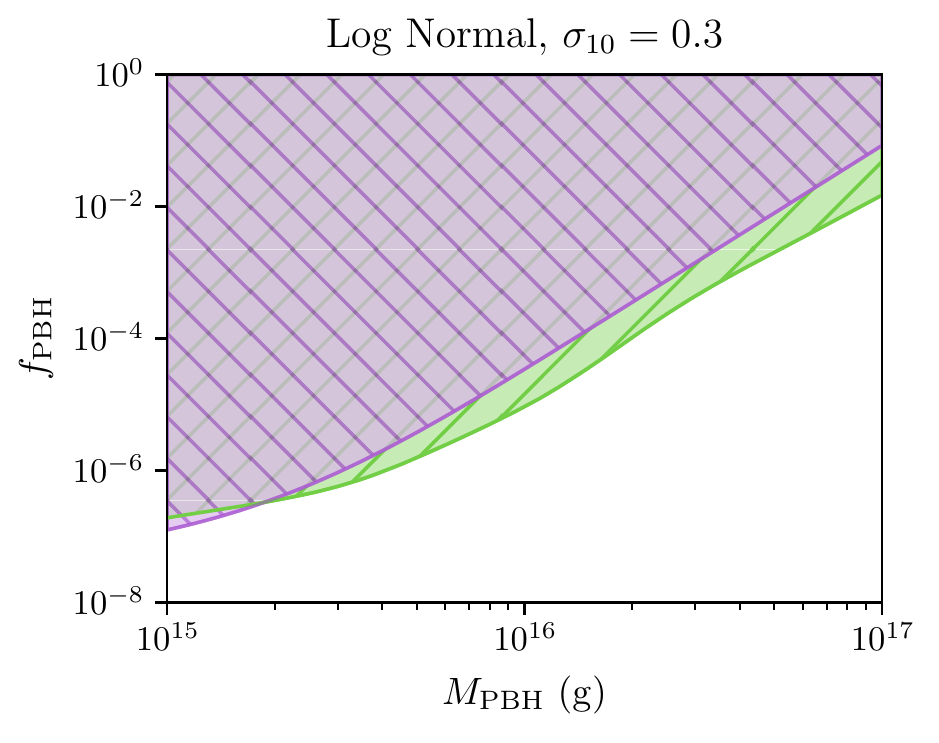}
    \label{subfig:gamma-cmb-lognorm}
    \vspace{-2em}
  \end{subfigure}
  \begin{subfigure}[t]{0.49\textwidth}
    \centering
    \includegraphics[height=2.4in]{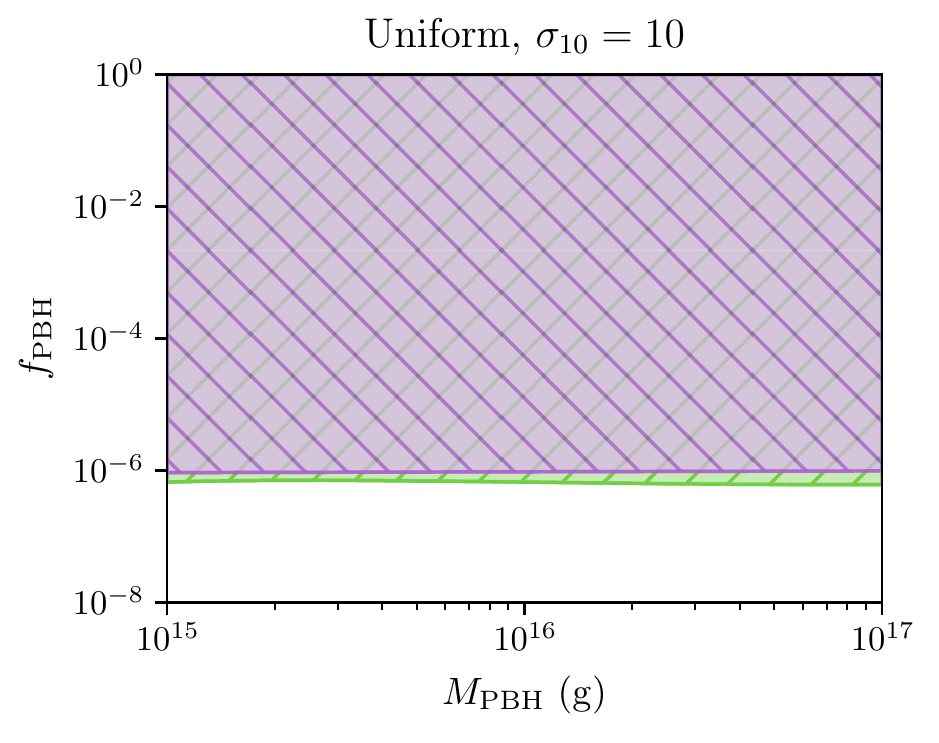}
    \label{subfig:gamma-cmb-uniform}
  \vspace{-2.0em}
  \end{subfigure}
  \caption{Comparison of \SI{95}{\percent} exclusion regions for evaporating PBHs derived from CMB and gamma ray background measurements. Here, base \LCDM parameters for the CMB regions are fixed to the values given in Table~\ref{tab:lcdm-parameters}, and gamma ray constraints are computed using the method for extending mass distributions given in Ref.~\cite{Carr:2017jsz}.}
  \label{fig:gamma-cmb-comp}
\end{figure}

We can also compare these limits with recent constraints from gamma-ray background measurements.
The constraint for this measurement in Eq.~\eqref{eq:pbh-frac-gamma-mono} is used to calculate the log-normal PBH fraction \(f_{\PBH,\text{log norm}}\) for different values of \(\sigma_\ten\).
These are then compared to the constraints derived from the full CMB analysis and are presented in Figure~\ref{fig:gamma-cmb-comp}.

We compare CMB results calculated with fixed base \LCDM parameters, as the gamma ray background constraints are calculated using fixed Hubble parameter and cold dark matter density values.
It would of course also be interesting, but beyond the scope of the work, to perform a re-analysis of the gamma-ray background constraints with varying \LCDM parameters.

What we see from examining Fig.~\ref{fig:gamma-cmb-comp} is that the constraints from both CMB and gamma ray background measurements remain close across both the log-normal and uniform distributions.
In the log-normal case, gamma ray background measurements are more constraining at low masses compared to CMB measurements, yet for intermediate masses \(M_\PBH \sim \SI{e16}{\gram}\), CMB constraints are stronger on \(f_\PBH\) by an order of magnitude.
This could be because mid-range PBHs do not eject as many gamma rays as their lighter counterparts, leading to slightly relaxed bounds for these masses.
On the other hand, CMB constraints are predominantly influenced by the total energy injected into the primordial plasma, which is much more for a log-normal than a monochromatic distribution, as we have already seen.

In the case of the uniform distribution, we see both constraints are practically indistinguishable.
However, we found in the previous section for a uniform distribution that the maximum allowed value of \(f_\PBH\) by extending the monochromatic mass constraints is larger than the computed value by a factor of roughly 4.5.
With the addition of this factor, the maximum allowed \(f_\PBH\) for the gamma ray constraint reduces to \(f_\PBH \sim \num{2.1e-7}\), making this constraint the dominant one for the uniform mass distribution.
Interesting to note is the effects of the truncation correction on the gamma ray constraints for the uniform distribution --- without this correction, the constraint sits at about \(f_\PBH \sim \num{1e-5}\), over an order of magnitude larger than the CMB constraint.

\section{Conclusions \label{sec:conclusions}}

Primordial black holes are an interesting candidate for dark matter with a rich phenomenology.
In this paper we have considered light PBHs in the mass range of \SIrange{e15}{e17}{\gram} injecting energy into the primordial plasma through Hawking radiation.

We have found for a monochromatic mass distribution that the fraction of dark matter made of PBHs \(f_\PBH\) shares slight degeneracies with the Hubble parameter \(H_0\) and the reionisation optical depth \(\tau\), where the latter is seen to be a ``true'' degeneracy when examining \(f_\PBH\) and \(\tau\) in isolation.
We also found that allowing the base \LCDM parameters to vary weakens bounds on \(f_\PBH\) by roughly an order of magnitude across the mass range, with constraints fully relaxed for PBHs with mass \(M_\PBH \gtrsim \SI{6e16}{\gram}\).
By examining constraints placed on the width of the log-normal distribution, \(\sigma_\ten\), we saw that decreasing \(f_\PBH\) lessens constraints on the mean mass much more than \(\sigma_\ten\), up to some critical point where the exclusion region turns convex.
Alongside this, we found that allowing \LCDM parameters to vary removes any constraints on the width of a log-normal distribution when \(f_\PBH = \num{e-7}\).
We showed that a uniform mass distribution is much more heavily constrained than a delta mass distribution, with \SI{95}{\percent} exclusion limits of \(f_\PBH < \num{6.7e-7}\) for fixed \LCDM base parameters and \(f_\PBH < \num{1.6e-5}\) for free \LCDM base parameters.
We also saw that a monochromatic mass distribution constrains light PBHs better than distributions with finite width, whilst for mean masses of \(\gtrsim \SI{3e15}{\gram}\) the bounds on \(f_\PBH\) are relaxed by a few orders of magnitude.

We examined the method presented in Ref.~\cite{Carr:2017jsz} which shows how to use constraints on monochromatic mass distributions for PBHs to compute constraints for extended mass distributions.
The comparison was made for both a log-normal and uniform mass distribution and showed excellent agreement in the case of the log-normal, with a slight overestimation by a factor of 4.5 for the uniform distribution.
This technique was then used to extend the latest competing gamma ray background constraints on PBHs, where we found that for both a log-normal and uniform distribution, the measurements constrain similar regions of parameter space.
Perhaps more interestingly, we saw for a log-normal that the mid- to high-mass PBH range is better constrained by CMB measurements, whilst the low-mass region is better constrained by gamma ray measurements.

\section*{Acknowledgments}
JH gratefully acknowledges the support of an Australian Research Council Future Fellowship (FT140100481).
HP acknowledges the support of a Masters (No Honours) scholarship from The University of Adelaide, and the use of the Phoenix supercomputing cluster.
MW is supported by the Australian Research Council Future Fellowship FT140100244.
YAH is supported by the National Science Foundation under grant No.~1820861.
MW wishes to thank Lucien Boland and Sean Crosby for their administration of, and ongoing assistance with, the MPI-enabled computing cluster on which this work was performed.
AGW is supported by the ARC Centre of Excellence for Particle Physics at the Terascale (CoEPP) (CE110001104) and the Centre for the Subatomic Structure of Matter (CSSM).

\bibliographystyle{JHEP}
\bibliography{references.bib}

\providecommand{\href}[2]{#2}\begingroup\raggedright\begin{thebibliography}{10}

\bibitem{Chapline:1975}
G.~F. {Chapline}, \emph{{Cosmological effects of primordial black holes}},
  \href{http://dx.doi.org/10.1038/253251a0}{\emph{Nature} {\bfseries 253}
  (Jan., 1975) 251--252}.

\bibitem{Hawking:1971ei}
S.~Hawking, \emph{{Gravitationally collapsed objects of very low mass}},
  {\emph{Mon. Not. Roy. Astron. Soc.} {\bfseries 152} (1971) 75}.

\bibitem{Carr:2016drx}
B.~Carr, F.~K{\"u}hnel and M.~Sandstad, \emph{{Primordial Black Holes as Dark
  Matter}}, \href{http://dx.doi.org/10.1103/PhysRevD.94.083504}{\emph{Phys.
  Rev.} {\bfseries D94} (2016) 083504},
  [\href{https://arxiv.org/abs/1607.06077}{{\ttfamily 1607.06077}}].

\bibitem{Niikura:2017zjd}
H.~Niikura, M.~Takada, N.~Yasuda, R.~H. Lupton, T.~Sumi, S.~More et~al.,
  \emph{{Microlensing constraints on primordial black holes with the Subaru/HSC
  Andromeda observation}},  \href{https://arxiv.org/abs/1701.02151}{{\ttfamily
  1701.02151}}.

\bibitem{Capela:2013yf}
F.~Capela, M.~Pshirkov and P.~Tinyakov, \emph{{Constraints on primordial black
  holes as dark matter candidates from capture by neutron stars}},
  \href{http://dx.doi.org/10.1103/PhysRevD.87.123524}{\emph{Phys. Rev.}
  {\bfseries D87} (2013) 123524},
  [\href{https://arxiv.org/abs/1301.4984}{{\ttfamily 1301.4984}}].

\bibitem{Afshordi:2003zb}
N.~Afshordi, P.~McDonald and D.~N. Spergel, \emph{{Primordial black holes as
  dark matter: The Power spectrum and evaporation of early structures}},
  \href{http://dx.doi.org/10.1086/378763}{\emph{Astrophys. J.} {\bfseries 594}
  (2003) L71--L74}, [\href{https://arxiv.org/abs/astro-ph/0302035}{{\ttfamily
  astro-ph/0302035}}].

\bibitem{Ricotti:2008}
M.~{Ricotti}, J.~P. {Ostriker} and K.~J. {Mack}, \emph{{Effect of Primordial
  Black Holes on the Cosmic Microwave Background and Cosmological Parameter
  Estimates}}, \href{http://dx.doi.org/10.1086/587831}{\emph{Astrophys. J.}
  {\bfseries 680} (June, 2008) 829--845},
  [\href{https://arxiv.org/abs/0709.0524}{{\ttfamily 0709.0524}}].

\bibitem{Carr:2009jm}
B.~J. Carr, K.~Kohri, Y.~Sendouda and J.~Yokoyama, \emph{{New cosmological
  constraints on primordial black holes}},
  \href{http://dx.doi.org/10.1103/PhysRevD.81.104019}{\emph{Phys. Rev.}
  {\bfseries D81} (2010) 104019},
  [\href{https://arxiv.org/abs/0912.5297}{{\ttfamily 0912.5297}}].

\bibitem{Bernal:2017vvn}
J.~Luis~Bernal, N.~Bellomo, A.~Raccanelli and L.~Verde, \emph{{Cosmological
  implications of Primordial Black Holes}},
  \href{http://dx.doi.org/10.1088/1475-7516/2017/10/052}{\emph{JCAP} {\bfseries
  1710} (2017) 052}, [\href{https://arxiv.org/abs/1709.07465}{{\ttfamily
  1709.07465}}].

\bibitem{AliHaimoud:2017}
Y.~{Ali-Ha{\"i}moud} and M.~{Kamionkowski}, \emph{{Cosmic microwave background
  limits on accreting primordial black holes}},
  \href{http://dx.doi.org/10.1103/PhysRevD.95.043534}{\emph{Phys. Rev.}
  {\bfseries 95} (Feb., 2017) 043534},
  [\href{https://arxiv.org/abs/1612.05644}{{\ttfamily 1612.05644}}].

\bibitem{Poulin:2017a}
V.~{Poulin}, P.~D. {Serpico}, F.~{Calore}, S.~{Clesse} and K.~{Kohri},
  \emph{{CMB bounds on disk-accreting massive primordial black holes}},
  \href{http://dx.doi.org/10.1103/PhysRevD.96.083524}{\emph{Phys. Rev.}
  {\bfseries 96} (Oct., 2017) 083524},
  [\href{https://arxiv.org/abs/1707.04206}{{\ttfamily 1707.04206}}].

\bibitem{Zhu:2017plg}
Q.~Zhu, E.~Vasiliev, Y.~Li and Y.~Jing, \emph{{Primordial Black Holes as Dark
  Matter: Constraints From Compact Ultra-Faint Dwarfs}},
  \href{http://dx.doi.org/10.1093/mnras/sty079}{\emph{Mon. Not. Roy. Astron.
  Soc.} {\bfseries 476} (2018) 2--11},
  [\href{https://arxiv.org/abs/1710.05032}{{\ttfamily 1710.05032}}].

\bibitem{AliHaimoud:2017b}
Y.~{Ali-Ha{\"i}moud}, E.~D. {Kovetz} and M.~{Kamionkowski}, \emph{{Merger rate
  of primordial black-hole binaries}},
  \href{http://dx.doi.org/10.1103/PhysRevD.96.123523}{\emph{\prd} {\bfseries
  96} (Dec., 2017) 123523}, [\href{https://arxiv.org/abs/1709.06576}{{\ttfamily
  1709.06576}}].

\bibitem{Katz:2018zrn}
A.~Katz, J.~Kopp, S.~Sibiryakov and W.~Xue, \emph{{Femtolensing by Dark Matter
  Revisited}},
  \href{http://dx.doi.org/10.1088/1475-7516/2018/12/005}{\emph{JCAP} {\bfseries
  1812} (2018) 005}, [\href{https://arxiv.org/abs/1807.11495}{{\ttfamily
  1807.11495}}].

\bibitem{Carr:2016hva}
B.~J. Carr, K.~Kohri, Y.~Sendouda and J.~Yokoyama, \emph{{Constraints on
  primordial black holes from the Galactic gamma-ray background}},
  \href{http://dx.doi.org/10.1103/PhysRevD.94.044029}{\emph{Phys. Rev.}
  {\bfseries D94} (2016) 044029},
  [\href{https://arxiv.org/abs/1604.05349}{{\ttfamily 1604.05349}}].

\bibitem{Fermi-LAT:2018pfs}
{\scshape Fermi-LAT} collaboration, M.~Ackermann et~al., \emph{{Search for
  Gamma-Ray Emission from Local Primordial Black Holes with the Fermi Large
  Area Telescope}},
  \href{http://dx.doi.org/10.3847/1538-4357/aaac7b}{\emph{Astrophys. J.}
  {\bfseries 857} (2018) 49},
  [\href{https://arxiv.org/abs/1802.00100}{{\ttfamily 1802.00100}}].

\bibitem{Barnacka:2012bm}
A.~Barnacka, J.~F. Glicenstein and R.~Moderski, \emph{{New constraints on
  primordial black holes abundance from femtolensing of gamma-ray bursts}},
  \href{http://dx.doi.org/10.1103/PhysRevD.86.043001}{\emph{Phys. Rev.}
  {\bfseries D86} (2012) 043001},
  [\href{https://arxiv.org/abs/1204.2056}{{\ttfamily 1204.2056}}].

\bibitem{Poulin:2016anj}
V.~Poulin, J.~Lesgourgues and P.~D. Serpico, \emph{{Cosmological constraints on
  exotic injection of electromagnetic energy}},
  \href{http://dx.doi.org/10.1088/1475-7516/2017/03/043}{\emph{JCAP} {\bfseries
  1703} (2017) 043}, [\href{https://arxiv.org/abs/1610.10051}{{\ttfamily
  1610.10051}}].

\bibitem{Pi:2017gih}
S.~Pi, Y.-L. Zhang, Q.-G. Huang and M.~Sasaki, \emph{{Scalaron from
  $R^2$-gravity as a heavy field}},
  \href{http://dx.doi.org/10.1088/1475-7516/2018/05/042}{\emph{JCAP} {\bfseries
  1805} (2018) 042}, [\href{https://arxiv.org/abs/1712.09896}{{\ttfamily
  1712.09896}}].

\bibitem{Clesse:2015wea}
S.~Clesse and J.~Garc{\'i}a-Bellido, \emph{{Massive Primordial Black Holes from
  Hybrid Inflation as Dark Matter and the seeds of Galaxies}},
  \href{http://dx.doi.org/10.1103/PhysRevD.92.023524}{\emph{Phys. Rev.}
  {\bfseries D92} (2015) 023524},
  [\href{https://arxiv.org/abs/1501.07565}{{\ttfamily 1501.07565}}].

\bibitem{Carr:2017jsz}
B.~Carr, M.~Raidal, T.~Tenkanen, V.~Vaskonen and H.~Veerm{\"a}e,
  \emph{{Primordial black hole constraints for extended mass functions}},
  \href{http://dx.doi.org/10.1103/PhysRevD.96.023514}{\emph{Phys. Rev.}
  {\bfseries D96} (2017) 023514},
  [\href{https://arxiv.org/abs/1705.05567}{{\ttfamily 1705.05567}}].

\bibitem{Kuhnel:2017pwq}
F.~K{\"u}hnel and K.~Freese, \emph{{Constraints on Primordial Black Holes with
  Extended Mass Functions}},
  \href{http://dx.doi.org/10.1103/PhysRevD.95.083508}{\emph{Phys. Rev.}
  {\bfseries D95} (2017) 083508},
  [\href{https://arxiv.org/abs/1701.07223}{{\ttfamily 1701.07223}}].

\bibitem{Bellomo:2017zsr}
N.~Bellomo, J.~L. Bernal, A.~Raccanelli and L.~Verde, \emph{{Primordial Black
  Holes as Dark Matter: Converting Constraints from Monochromatic to Extended
  Mass Distributions}},
  \href{http://dx.doi.org/10.1088/1475-7516/2018/01/004}{\emph{JCAP} {\bfseries
  1801} (2018) 004}, [\href{https://arxiv.org/abs/1709.07467}{{\ttfamily
  1709.07467}}].

\bibitem{Belotsky:2014twa}
K.~M. Belotsky and A.~A. Kirillov, \emph{{Primordial black holes with mass
  $10^{16}-10^{17}$ g and reionization of the Universe}},
  \href{http://dx.doi.org/10.1088/1475-7516/2015/01/041}{\emph{JCAP} {\bfseries
  1501} (2015) 041}, [\href{https://arxiv.org/abs/1409.8601}{{\ttfamily
  1409.8601}}].

\bibitem{Clark:2016nst}
S.~Clark, B.~Dutta, Y.~Gao, L.~E. Strigari and S.~Watson, \emph{{Planck
  Constraint on Relic Primordial Black Holes}},
  \href{http://dx.doi.org/10.1103/PhysRevD.95.083006}{\emph{Phys. Rev.}
  {\bfseries D95} (2017) 083006},
  [\href{https://arxiv.org/abs/1612.07738}{{\ttfamily 1612.07738}}].

\bibitem{Page:1976ki}
D.~N. Page, \emph{{Particle Emission Rates from a Black Hole. 2. Massless
  Particles from a Rotating Hole}},
  \href{http://dx.doi.org/10.1103/PhysRevD.14.3260}{\emph{Phys. Rev.}
  {\bfseries D14} (1976) 3260--3273}.

\bibitem{Hawking:1974rv}
S.~W. Hawking, \emph{{Black hole explosions}},
  \href{http://dx.doi.org/10.1038/248030a0}{\emph{Nature} {\bfseries 248}
  (1974) 30--31}.

\bibitem{Hawking:1974sw}
S.~W. Hawking, \emph{{Particle Creation by Black Holes}},
  \href{http://dx.doi.org/10.1007/BF02345020, 10.1007/BF01608497}{\emph{Commun.
  Math. Phys.} {\bfseries 43} (1975) 199--220}.

\bibitem{MacGibbon:1990zk}
J.~H. MacGibbon and B.~R. Webber, \emph{{Quark and gluon jet emission from
  primordial black holes: The instantaneous spectra}},
  \href{http://dx.doi.org/10.1103/PhysRevD.41.3052}{\emph{Phys. Rev.}
  {\bfseries D41} (1990) 3052--3079}.

\bibitem{MacGibbon:1991tj}
J.~H. MacGibbon, \emph{{Quark and gluon jet emission from primordial black
  holes. 2. The Lifetime emission}},
  \href{http://dx.doi.org/10.1103/PhysRevD.44.376}{\emph{Phys. Rev.} {\bfseries
  D44} (1991) 376--392}.

\bibitem{Chluba_13}
J.~{Chluba}, \emph{{Green's function of the cosmological thermalization
  problem}}, \href{http://dx.doi.org/10.1093/mnras/stt1025}{\emph{Mon. Not.
  Roy. Astron. Soc.} {\bfseries 434} (Sept., 2013) 352--357},
  [\href{https://arxiv.org/abs/1304.6120}{{\ttfamily 1304.6120}}].

\bibitem{Peebles:1968ja}
P.~J.~E. Peebles, \emph{{Recombination of the Primeval Plasma}},
  \href{http://dx.doi.org/10.1086/149628}{\emph{Astrophys. J.} {\bfseries 153}
  (1968) 1}.

\bibitem{Zeldovich:1969en}
{\relax Ya}.~B. Zeldovich, V.~G. Kurt and R.~A. Sunyaev, \emph{{Recombination
  of hydrogen in the hot model of the universe}}, {\emph{Sov. Phys. JETP}
  {\bfseries 28} (1969) 146}.

\bibitem{Chen:2003gz}
X.-L. Chen and M.~Kamionkowski, \emph{{Particle decays during the cosmic dark
  ages}}, \href{http://dx.doi.org/10.1103/PhysRevD.70.043502}{\emph{Phys. Rev.}
  {\bfseries D70} (2004) 043502},
  [\href{https://arxiv.org/abs/astro-ph/0310473}{{\ttfamily
  astro-ph/0310473}}].

\bibitem{Slatyer:2015kla}
T.~R. Slatyer, \emph{{Indirect Dark Matter Signatures in the Cosmic Dark Ages
  II. Ionization, Heating and Photon Production from Arbitrary Energy
  Injections}}, \href{http://dx.doi.org/10.1103/PhysRevD.93.023521}{\emph{Phys.
  Rev.} {\bfseries D93} (2016) 023521},
  [\href{https://arxiv.org/abs/1506.03812}{{\ttfamily 1506.03812}}].

\bibitem{Green:2016xgy}
A.~M. Green, \emph{{Microlensing and dynamical constraints on primordial black
  hole dark matter with an extended mass function}},
  \href{http://dx.doi.org/10.1103/PhysRevD.94.063530}{\emph{Phys. Rev.}
  {\bfseries D94} (2016) 063530},
  [\href{https://arxiv.org/abs/1609.01143}{{\ttfamily 1609.01143}}].

\bibitem{Kannike:2017bxn}
K.~Kannike, L.~Marzola, M.~Raidal and H.~Veerm{\"a}e, \emph{{Single Field
  Double Inflation and Primordial Black Holes}},
  \href{http://dx.doi.org/10.1088/1475-7516/2017/09/020}{\emph{JCAP} {\bfseries
  1709} (2017) 020}, [\href{https://arxiv.org/abs/1705.06225}{{\ttfamily
  1705.06225}}].

\bibitem{Blas:2011rf}
D.~Blas, J.~Lesgourgues and T.~Tram, \emph{{The Cosmic Linear Anisotropy
  Solving System (CLASS) II: Approximation schemes}},
  \href{http://dx.doi.org/10.1088/1475-7516/2011/07/034}{\emph{JCAP} {\bfseries
  1107} (2011) 034}, [\href{https://arxiv.org/abs/1104.2933}{{\ttfamily
  1104.2933}}].

\bibitem{Seager:1999bc}
S.~Seager, D.~D. Sasselov and D.~Scott, \emph{{A new calculation of the
  recombination epoch}},
  \href{http://dx.doi.org/10.1086/312250}{\emph{Astrophys. J.} {\bfseries 523}
  (1999) L1--L5}, [\href{https://arxiv.org/abs/astro-ph/9909275}{{\ttfamily
  astro-ph/9909275}}].

\bibitem{AliHaimoud:2010dx}
Y.~Ali-Ha{\"i}moud and C.~M. Hirata, \emph{{HyRec: A fast and highly accurate
  primordial hydrogen and helium recombination code}},
  \href{http://dx.doi.org/10.1103/PhysRevD.83.043513}{\emph{Phys. Rev.}
  {\bfseries D83} (2011) 043513},
  [\href{https://arxiv.org/abs/1011.3758}{{\ttfamily 1011.3758}}].

\bibitem{AliHaimoud:2010ab}
Y.~Ali-Ha{\"i}moud and C.~M. Hirata, \emph{{Ultrafast effective multi-level
  atom method for primordial hydrogen recombination}},
  \href{http://dx.doi.org/10.1103/PhysRevD.82.063521}{\emph{Phys. Rev.}
  {\bfseries D82} (2010) 063521},
  [\href{https://arxiv.org/abs/1006.1355}{{\ttfamily 1006.1355}}].

\bibitem{Giesen:2012}
G.~{Giesen}, J.~{Lesgourgues}, B.~{Audren} and Y.~{Ali-Ha{\"i}moud}, \emph{{CMB
  photons shedding light on dark matter}},
  \href{http://dx.doi.org/10.1088/1475-7516/2012/12/008}{\emph{JCAP} {\bfseries
  12} (Dec., 2012) 008}, [\href{https://arxiv.org/abs/1209.0247}{{\ttfamily
  1209.0247}}].

\bibitem{Aghanim:2015xee}
{\scshape Planck} collaboration, N.~Aghanim et~al., \emph{{Planck 2015 results.
  XI. CMB power spectra, likelihoods, and robustness of parameters}},
  \href{http://dx.doi.org/10.1051/0004-6361/201526926}{\emph{Astron.
  Astrophys.} {\bfseries 594} (2016) A11},
  [\href{https://arxiv.org/abs/1507.02704}{{\ttfamily 1507.02704}}].

\bibitem{Feroz:2013hea}
F.~Feroz, M.~P. Hobson, E.~Cameron and A.~N. Pettitt, \emph{{Importance Nested
  Sampling and the MultiNest Algorithm}},
  \href{https://arxiv.org/abs/1306.2144}{{\ttfamily 1306.2144}}.

\bibitem{Feroz:2007kg}
F.~Feroz and M.~P. Hobson, \emph{{Multimodal nested sampling: an efficient and
  robust alternative to MCMC methods for astronomical data analysis}},
  \href{http://dx.doi.org/10.1111/j.1365-2966.2007.12353.x}{\emph{Mon. Not.
  Roy. Astron. Soc.} {\bfseries 384} (2008) 449},
  [\href{https://arxiv.org/abs/0704.3704}{{\ttfamily 0704.3704}}].

\bibitem{Feroz:2008xx}
F.~Feroz, M.~P. Hobson and M.~Bridges, \emph{{MultiNest: an efficient and
  robust Bayesian inference tool for cosmology and particle physics}},
  \href{http://dx.doi.org/10.1111/j.1365-2966.2009.14548.x}{\emph{Mon. Not.
  Roy. Astron. Soc.} {\bfseries 398} (2009) 1601--1614},
  [\href{https://arxiv.org/abs/0809.3437}{{\ttfamily 0809.3437}}].

\bibitem{Riess:2016jrr}
A.~G. Riess et~al., \emph{{A 2.4\% Determination of the Local Value of the
  Hubble Constant}},
  \href{http://dx.doi.org/10.3847/0004-637X/826/1/56}{\emph{Astrophys. J.}
  {\bfseries 826} (2016) 56},
  [\href{https://arxiv.org/abs/1604.01424}{{\ttfamily 1604.01424}}].

\bibitem{Aghanim:2018eyx}
{\scshape Planck} collaboration, N.~Aghanim et~al., \emph{{Planck 2018 results.
  VI. Cosmological parameters}},
  \href{https://arxiv.org/abs/1807.06209}{{\ttfamily 1807.06209}}.

\end{thebibliography}\endgroup

\end{document}